\documentclass[10pt]{article}
\usepackage{graphicx}
\usepackage{amssymb}
\usepackage{amscd}
\usepackage{amsmath}
\usepackage[english]{babel}
\usepackage{axodraw}
\usepackage[T2A]{fontenc}
\usepackage{EuScript}
\usepackage{eufrak}

\def\be{\begin{eqnarray}}
\def\ee{\end{eqnarray}}

\def\p{\partial}

\newcommand{\labcd}[2]{\hbox to\textwidth{#1\dotfill #2}}

\hoffset= - 1.10in         

\begin{document}

\hfill ITEP/TH - 08/06

\hfill INR/TH - 06-2006

\bigskip

\centerline{\Large{{\bf Quantum Field Theory as Effective BV Theory
from Chern-Simons}}}

\bigskip
\centerline{Dmitry Krotov$^{a}$ and Andrei Losev$^{b}$}

\begin{center}
$^a${\small{\em
Institute for Nuclear Research of the Russian Academy of Sciences, }}\\
{\small{\em
60th October Anniversary prospect 7a, Moscow, 117312, Russia
}}\\
$^{a,b}${\small{\em
Institute of Theoretical and Experimental Physics, }}\\
{\small{\em
B.Cheremushkinskaya 25, Moscow, 117259, Russia
}}\\
$^a${\small{\em
Moscow State University, Department of Physics,}}\\
{\small{\em
Vorobjevy Gory, Moscow, 119899, Russia
}}
\end{center}
\bigskip

\centerline{ABSTRACT}

\bigskip

 The general procedure for obtaining explicit expressions for
all cohomologies of N.Berkovits's operator  is suggested. It is
demonstrated that calculation of BV integral for the classical
Chern-Simons-like theory (Witten's OSFT-like theory) reproduces BV
version of two dimensional gauge model at the level of effective
action. This model contains gauge field, scalars, fermions and some
other fields. We prove that this model is an example of "{\bf
singular}" point from the perspective of the suggested method for
cohomology evaluation. For arbitrary "{\bf regular}" point the same
technique results in AKSZ(Alexandrov, Kontsevich, Schwarz,
Zaboronsky) version of Chern-Simons theory (BF theory) in accord
with \cite{Berkovits,Movshev}.

\bigskip

\section{Introduction}
 The search for fundamental degrees of freedom is a challenging problem
in physics.  Different attempts in this direction demonstrated that
Chern-Simons-like theories (Witten's OSFT-like theories) play
important role for questions of this type \cite{Witten}. In the
present paper we discuss a small subject in the story which to the
best of our knowledge did not attract enough attention in the
literature.  We suggest the construction which allows to obtain a
classical action for a large class of quantum field theories  as
effective Batalin-Vilkovisky(BV) action from  the  certain
fundamental action (\ref{Fund}). This realizes a beautiful idea that
different "physically interesting" theories can arise in a universal
way as effective actions from the certain fundamental theory. The
technique is very much in the spirit of recently suggested formalism
for covariant quantization of the superparticle \cite{Berkovits} and
the superstring \cite{BerkovitsSuperString} by N.Berkovits. Close
ideas were discussed in \cite{Movshev}-\cite{Lysov} and in
\cite{Grigoriev}.

  The basis of construction is the space (algebra)
$O[\theta_\alpha\ \! ,\ \! \lambda_\alpha\ |\ f^{\mu}(\lambda)]$
which is generated by the set of odd variables $\theta_\alpha$, even
variables $\lambda_\alpha$ modulo the system of constraints
$f^{\mu}(\lambda)$, quadratic in $\lambda_\alpha$ ($\alpha=1...K$,
$\mu=1...N$). This  means that if some function of $\lambda$ and
$\theta$ is multiplied  by $f^{\mu}(\lambda)$,  such product is
equivalent to zero in the space $O[\theta_\alpha\ \! ,\ \!
\lambda_\alpha\ |\ f^{\mu}(\lambda)]$. In other words, one should
consider the supercommutative ring $C[\lambda,\ \theta]$  of
polynomials in $\lambda_\alpha$ and $\theta_\alpha$. This ring
contains the ideal $I_f$ which is generated by the set of quadratic
constraints $f^{\mu}(\lambda)$, i.e. arbitrary element in $I_f$ can
be written as $c_{\mu}(\lambda,\theta)f^{\mu}(\lambda)$ where
coefficients $c_{\mu}(\lambda,\theta)\in C[\lambda,\ \theta]$ belong
to the ring $C[\lambda,\theta]$. The space $O[\theta_\alpha\ \! ,\
\! \lambda_\alpha\ |\ f^{\mu}(\lambda)]$ is given by the coset
$O[\theta_\alpha\ \! ,\ \! \lambda_\alpha\ |\ f^{\mu}(\lambda)]=
C[\lambda,\ \theta]\diagup I_f$.

  In the recent discussions
\cite{Berkovits,BerkovitsSuperString,Pure Spinors} notation
$\lambda_\alpha$ was used for pure spinor variables which satisfy
pure spinor constraints $f^\mu(\lambda)\ =\
\lambda_\alpha\gamma^{\mu}_{\alpha\beta}\lambda_\beta$. One message
of the present paper is to relax this condition and to consider
arbitrary system of constraints $f^{\mu}(\lambda)$, quadratic in
$\lambda_\alpha$, which are not necessarily connected to $\gamma$ -
matrices. It will be demonstrated that this setup leads to
non-trivial effective gauge model at least in two dimensions.\ Thus,
since  $\lambda_\alpha$  are no longer pure  spinors,  we call  them
simply  even variables and functions $f^\mu(\lambda)$ which
generalize pure spinor constraints  - simply constraints (generating
the ideal  $I_f$).

  The construction starts from the fundamental action written as
\begin{equation}\label{Fund}
S^{Fund}\ =\ \int\ STr\Big( <\EuScript{P},\  Q_{B} \EuScript{A}>\ +\
g<\EuScript{P},\ \EuScript{A}^2>\Big)
\end{equation}
where the field  $\EuScript{A}$ belongs to space $O[\theta_\alpha\
\! ,\ \! \lambda_\alpha\ |\ f^{\mu}(\lambda)]\otimes Func(x)\otimes
T(\mathfrak{g})$. $Func(x)$ is  a space of functions of space-time
coordinates and $T(\mathfrak{g})$ is a representation of semi-simple
gauge algebra $\mathfrak{g}$. The field $\EuScript{P}$ belongs to
the dual space $O[\theta_\alpha\ \! ,\ \! \lambda_\alpha\ |\
f^{\mu}(\lambda)]^\ast\otimes Func(x)\otimes T(\mathfrak{g})$. Thus,
each field belongs to space which is the tensor product of three
spaces. To construct a scalar action from these fields introduce
three pairings: integration - the pairing in the space $Func(x)$,
$STr$ - the supertrace - the pairing in $T(\mathfrak{g})$ and $<\ ,\
>$ - the canonical paring\footnote{\label{dual} It should be
emphasized that in the main text of the paper we denote the elements
of the dual basis in the space $O[\theta_\alpha\ \! ,\ \!
\lambda_\alpha\ |\ f^{\mu}(\lambda)]^\ast$ by the same symbols as
the elements of  the basis in the space $O[\theta_\alpha\ \! ,\ \!
\lambda_\alpha\ |\ f^{\mu}(\lambda)]$ but with the underline sign.
For example, if $\lambda_1\theta_1$ is the basis element in the
space $O[\theta_\alpha\ \! ,\ \! \lambda_\alpha\ |\
f^{\mu}(\lambda)]$, then the dual basis element in $O[\theta_\alpha\
\! ,\ \! \lambda_\alpha\ |\ f^{\mu}(\lambda)]^\ast$ would be denoted
by $\underline{\lambda_1\theta_1}$. The canonical pairing between
these two elements is $<\underline{\lambda_1\theta_1}\ ,\
\lambda_1\theta_1>\ =\ 1$. The basis elements in the ring
$O[\theta_\alpha\ \! ,\ \! \lambda_\alpha\ |\ f^{\mu}(\lambda)]$ are
given by all possible monoms in $\lambda_\alpha$ and
$\theta_\alpha$.} between the elements of the space
$O[\theta_\alpha\ \! ,\ \! \lambda_\alpha\ |\ f^{\mu}(\lambda)]$ and
the elements of $O[\theta_\alpha\ \! ,\ \! \lambda_\alpha\ |\
f^{\mu}(\lambda)]^\ast$. The definition of the supertrace is given
in the footnote \ref{supertrace}. Parameter $g$ in the action
(\ref{Fund}) is  a coupling constant.

The central object of our consideration is the operator
\cite{Berkovits,BerkovitsSuperString,Witten and Nilsson}:
\begin{equation}\label{Qtot}
Q_{B}\ =\ Q\ +\ \Phi\ =\
\lambda_\alpha\frac{\partial}{\partial\theta_\alpha}\ +\
\theta_{\alpha}\frac{\partial
f^{\mu}}{\partial\lambda_\alpha}\partial_{\mu}
\end{equation}
Here $\partial_\mu=\frac{\partial}{\partial x^{\mu}}$ is a
derivative with respect to the space-time coordinates. Operator
$Q_B$ is nilpotent ($Q_B^2\ =\ 0$) in the space $O[\theta_\alpha\ \!
,\ \! \lambda_\alpha\ |\ f^{\mu}(\lambda)]$ due to  constraints
$f^{\mu}(\lambda)$.

In \cite{Berkovits,BerkovitsSuperString} it was suggested to extract
dynamical fields of the theory from the cohomologies of $Q\ =\
\lambda_\alpha\frac{\partial}{\partial\theta_\alpha}$ operator (\
the first term in (\ref{Qtot})\ ). These cohomologies
$\mathcal{H}(Q,\ O)$ or simply $\mathcal{H}(Q)$ should be calculated
in the space $O[\theta_\alpha\ \! ,\ \! \lambda_\alpha\ |\
f^{\mu}(\lambda)]$ which is given by the coset over the ideal $I_f$.
Because of this factorization calculation of cohomologies is rather
involved mathematical problem. It was attacked in \cite{Nekrasov}.
In this paper effective method of localization was used to compute
them. Such technique is very powerful in symmetric case (when the
matrices $\gamma^\mu_{\alpha \beta}$ in the definition of quadrics
$f^\mu(\lambda)\ = \ \lambda_\alpha\gamma^\mu_{\alpha
\beta}\lambda_\beta$ are governed by some large group  of symmetry).

  In  the present paper we suggest the general formulas which allow
to find explicit expressions for all representatives of
$\mathcal{H}(Q)$  for arbitrary system of quadrics $f^\mu(\lambda)$
(both in symmetric and non-symmetric cases). This calculation works
in a universal way independently of the particular structure of
functions $f^\mu(\lambda)$, their number $N$ ($\mu\ =\ 1...N$) and
the number $K$ of $\lambda_\alpha$ and $\theta_\alpha$ ($\alpha\ =\
1...K$). The description of this calculation is presented in the
chapter 4 and in the appendix.
\begin{figure}[h]
\centerline{\includegraphics[width=70mm,angle=-90]{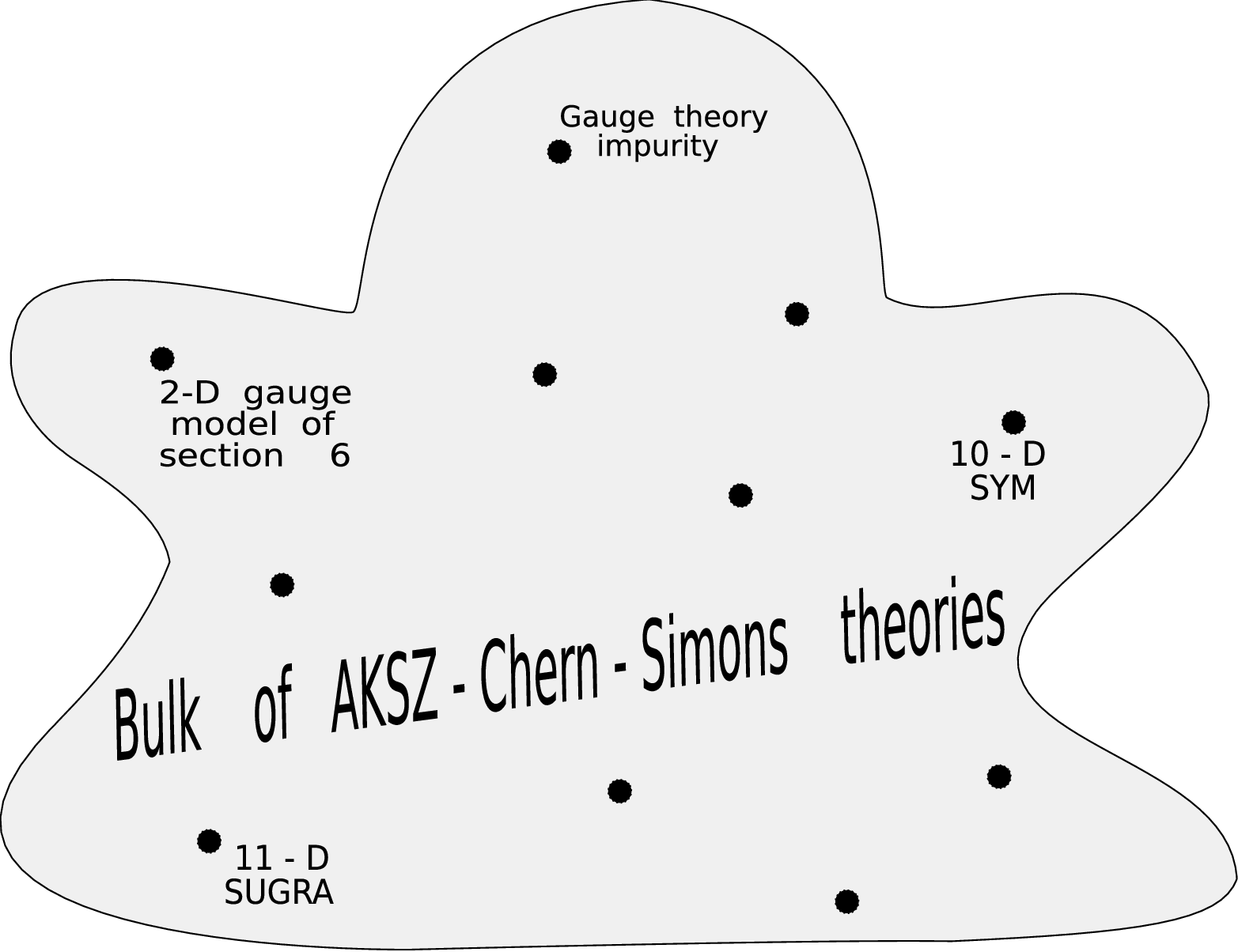}}
\caption{{\footnotesize The space of effective theories. Black dots
denote different gauge theories which are a kind of "singularities"
in the bulk of AKSZ-Chern-Simons theories, which are denoted by gray
background. This picture is in fact oversimplified and a little bit
cheating. The more accurate one is given in the conclusion (see
figure \ref{conclusion}).}} \label{landscape}
\end{figure}

 Another message of the present paper is to realize the
construction of \cite{Berkovits,BerkovitsSuperString} as calculation
of effective action for the certain fundamental action (\ref{Fund}).
Close attempts were made in \cite{Movshev,Prosto Movshev}. The
construction is given by the following steps. Firstly, using the
powerful method of chapter 4 one should find explicit expressions
for all representatives of $\mathcal{H}(Q)$. Then one should apply
the background field technique to find effective action for
(\ref{Fund}) taking as external (background) field the
representatives of $\mathcal{H}(Q)$. Effective action found in such
a way reproduces BV version of some "physically interesting" theory.

  According to this construction the whole information about degrees
of freedom and the structure of effective theory is encoded in the
representatives of $\mathcal{H}(Q)$, hence - in the system of
quadrics $f^\mu(\lambda)$. From the perspective of the method of
chapter 4 for cohomology evaluation there are two classes of
quadrics $f^\mu(\lambda)$. They are: "{\bf regular}"
system\footnote{In mathematical literature such systems are called
Koszul regular.} and "{\bf singular}" system. For arbitrary "{\bf
regular}" system of quadrics effective action is BV version of
AKSZ-Chern-Simons theory. In section 5 we note that nearly all
possible systems of quadrics $f^\mu(\lambda)$ are "{\bf regular}".
Still, there exists a small (but infinite) number of "{\bf
singular}" systems of quadrics. For such quadrics effective action
is BV  version of a non-trivial theory. One example of such "{\bf
singular}" point is the two-dimensional  gauge model of section 6.
Another one is 10D SYM theory of \cite{Berkovits,Movshev} and
probably even the supergravity theory of \cite{SUGRA}, see also
\cite{Vanhove}. This allows to think about "non-trivial" theories,
like 10D SYM and super-gauge model of section 6, as a kind of "{\bf
singularities}" in the space of "trivial" theories of Chern-Simons
type. This idea is illustrated in the figure \ref{landscape}. In
this picture the bulk of trivial theories like Chern-Simons is
denoted by gray background, while physically interesting gauge
theories are denoted by black dots.

 The universal feature of all these theories is that they are all
supersymmetric. However in {\bf "regular"} case this supersymmetry
transformation acts by zero (see \cite{we} for details).\\

\noindent  Summarizing the introduction, we emphasize that the main
new results of the present paper are:
\begin{enumerate}
\item Explicit expressions (\ref{coh}) for all representatives of $\mathcal{H}(Q,\
O)$ for arbitrary system of quadratic constraints $f^\mu(\lambda)$.

\item The two-dimensional  gauge model of section 6
can be obtained as effective action from the theory (\ref{Fund}).
The classical action for this theory is given in (\ref{intr
classical action}). In the spectrum of this theory there are: gauge
field (which is not dynamical), left and right fermions, adjoint
scalars $\phi_1$ and $\phi_2$ and some other fields. The system of
quadrics for this model is given by (\ref{introduction singular
constraints}).
\end{enumerate}

 The quadrics (\ref{introduction singular constraints}) is the
minimal system (minimal number of $\theta_\alpha$, $K\ =\ 4$ and
minimal number of quadrics, $N\ =\ 5$) which reproduces kinetic term
quadratic in derivatives in the effective action. It happens that
this particular model is two-dimensional (d=2). It seems probable,
that increasing the number of $\theta_\alpha$ ($K\ =\ 5,6,...$) and
the number of quadrics ($N\ =\ 6,7,...$) one can obtain higher
dimensional (d\ =\ 3,4,5,...) "physically interesting" theories in
the "{\bf singular}" points additionally to the known ones (10-D SYM
and 11-D SUGRA). To our mind it is of interest to search for such
"{\bf singular}" points in the space of quadrics $f^\mu(\lambda)$.
It is also of interest to study the string version of the model
suggested in section 6.

\subsection{Organization of the paper}
Below we give the outline of each chapter of the paper. For those
people who is interested in calculation of cohomologies
$\mathcal{H}(Q, O)$ we recommend to read directly chapter 4. While
our main result in this paper - the 2-D gauge model - is presented
in chapter 6.

\bigskip

\noindent {\bf 2. AKSZ Theory as BV Theory}
..................................................................................................................{\bf
4}

 This chapter does not contain new results. It is a kind of
introduction in some elementary facts about BV formalism. We give
the definition of AKSZ-theory and explain why we are interested in
this class of theories. Important statement is that the calculation
of BV integral for the classical BV action reproduces effective
action, which again satisfies BV Master Equation. It is explained
that quantum calculations with the Faddeev-Popov action are
equivalent to integration of BV action over the lagrangian
sub-manifold. This technique is used in chapter 3.

\bigskip

\noindent{\bf 3. Quantum Calculations}
..............................................................................................................................{\bf
8}

  It is explained how one should conduct Feynman diagram
calculations for the theory (\ref{Fund}). This is non-trivial since
the action (\ref{Fund}) has a large gauge freedom, hence requires
the choice of certain lagrangian sub-manifold. In our calculations
this is equivalent to particular choice of gauge in the theory
(\ref{Fund}).

\bigskip

\noindent{\bf 4. Calculation of Cohomologies}
................................................................................................................{\bf
10}

  This is the central chapter of the paper. It contains the new
result - explicit expressions (\ref{coh}) for all cohomologies of
$Q$-operator in the space $O[\theta_\alpha\ \! ,\ \! \lambda_\alpha\
|\ f^{\mu}(\lambda)]$\ (these cohomologies are usually called the
zero-momentum cohomologies). Subsection 4.1 contains the description
of the procedure, subsection 4.2 the brute force proof of this
result. Another, more elegant proof is presented in the appendix. In
the end of chapter 4 the definition of "{\bf regular}" and "{\bf
singular}" system of quadrics is given.

\bigskip

\noindent{\bf 5. AKSZ-Chern-Simons as Effective BV Theory}
....................................................................................{\bf
15}

In this chapter we present the calculation of effective BV action
for the simplest possible example - 3-dimensional AKSZ version of
Chern-Simons theory. The system of quadrics for this theory is given
by:
\begin{eqnarray}\label{introduction quadrics for CS }
f_{1}=\lambda_1^{2}\nonumber \\
f_{2}=\lambda_2^{2} \\
f_{3}=\lambda_3^{2}\nonumber
\end{eqnarray}
Though the answer for the cohomologies is almost obvious in this
case, in section 5.1 we apply the general procedure of chapter 4 to
find them. The aim of this calculation is to illustrate how the
procedure works in this simplest case.

 Then, in the subsection 5.2 we present the calculation of the
4-point amplitude in the theory (\ref{Fund}), emphasizing all
subtleties that are important for Feynman diagram calculations in
such theories. At this step this calculation is  just an
illustration, clarifying the conventions we use. As we explain such
amplitude is forbidden for the {\bf "regular"} set
(\ref{introduction quadrics for CS }). However, exactly this
amplitude will be important in calculation of effective action for
our main example  - the 2-dimensional gauge model of section 6.

Subsection 5.3 contains the calculation of effective action for the
system (\ref{introduction quadrics for CS }). The result is given in
(\ref{EFF-PCS}). It is demonstrated that (\ref{EFF-PCS}) reproduces
BV version of AKSZ-Chern-Simons theory from section 2.

  In the subsection 5.4 it is proved that the same result
(\ref{EFF-PCS}) for the effective action is true for arbitrary "{\bf
regular}" system of quadrics, not necessarily restricted to the
simple form (\ref{introduction quadrics for CS }). It is noted that
nearly all possible systems of quadrics $f^\mu(\lambda)$ are "{\bf
regular}".

\bigskip

\noindent{\bf 6. The Gauge Model}
....................................................................................................................................{\bf
20}

This chapter contains our main result in the present paper. We
suggest the example of "{\bf singular}" system of quadrics:
\begin{eqnarray}\label{introduction singular constraints}
f_1\ =\ \lambda_1\lambda_2\nonumber\\
f_2\ =\ \lambda_2\lambda_3\nonumber\\
f_3\ =\ \lambda_3\lambda_4\\
f_4\ \ \ =\ \ \lambda_1^2\nonumber\\
f_5\ \ \ =\ \ \lambda_4^2\nonumber
\end{eqnarray}
which under reduction over 3 dimensions reproduces BV action of
2-dimensional gauge model. The classical action, corresponding to
this BV action is given by
\begin{equation}\label{intr classical action}
\begin{split}
I[A]\ =\ \int\ Tr\ \bigg(\ \Phi F_{+ -}\ +\ D_+\phi_1D_-\phi_1\ +\
D_-\phi_2D_+\phi_2\ -\ \frac{g}{\sqrt{2}}\phi_1\{\psi_+,\psi_-\}\ +\
i\frac{g}{\sqrt{2}}\phi_2\{\psi_+,\psi_-\}\ +\ \beta_+D_-\gamma_+\
+\\ +\ \beta_-D_+\gamma_-\ +\ \overline{\psi}_-D_+\psi_-\ +\
\overline{\psi}_+D_-\psi_+\ +\ \overline{\chi}_-D_+\chi_-\ +\
\overline{\chi}_+D_-\chi_+\ +\ 2g\overline{\chi}_-[\gamma_-,\psi_+]\
+\ 2g\overline{\chi}_+[\gamma_+,\psi_-]\ \bigg)
\end{split}
\end{equation}
This model is another example of "physically interesting" theory
which like 10-dimensional SYM arises from Berkovits construction for
the "{\bf singular}" systems of quadrics $f^\mu(\lambda)$. However,
in comparison with 10-d SYM theory, this model is much  simpler,
hence is  more calculable.

\bigskip

\noindent{\bf 8. Appendix. Algebraic Meaning of the Berkovits
Complex}..............................................................{\bf
27}

 Another proof of the theorem from the section 4 is given. It is
demonstrated that Berkovits complex naturally arises for the pure
algebraic reasons.

\section{AKSZ theory as BV theory}\label{description of BV}
In this section we explain how AKSZ theory (definition is in the end
of the section) arises from the general BV formalism for gauge
theories \cite{BV,AS}. Consider a set of dynamical fields $\chi^n$
(arbitrary fields in our theory). For each field $\chi^n$, introduce
an anti-field $\chi_n^{\ast}$ with the opposite statistics to
$\chi^n$. Define the odd Laplace operator (BV laplacian) as
$\Delta_{BV}\ =\ \int
\frac{\delta_L}{\delta\chi^n(x)}\frac{\delta_R}{\delta
\chi_n^{\ast}(x)}$. Here sub-scripts $L$ and $R$ stand for left and
right derivatives. The central object of our consideration is BV
action, which is defined as a solution of the Master Equation:
\begin{equation}\label{quantum BV equation}
\triangle_{BV}\ e^{-\frac{1}{\hbar}
S^{BV}(\chi,\chi^{\ast})}\ =\ 0
\end{equation}
This equation in the limit $\hbar\rightarrow 0$ can be rewritten as
\begin{equation}\label{classical BV equation}
\int \frac{\ \ \delta_L S^{BV}}{\delta\chi^n(x)}\  \frac{\ \
\delta_R S^{BV}}{\delta\chi_n^{\ast}(x)}\ =\ 0
\end{equation}
which is called the classical  Master Equation.

  Quantum field theory in BV space is defined as an integral over
$\chi^n$ and $\chi_n^{\ast}$, restricted to lagrangian
sub-manifold\footnote{\label{Lagrang subman definition}Lagrangian
sub-manifold is usually defined as a sub-manifold satisfying
$\delta\chi_{n}^{\ast}\wedge \delta\chi^{n} \big|_{\EuScript{L}}\ =\
0$, where $\delta$ is de Rham operator on the space of fields. Each
lagrangian sub-manifold can be locally described by its generating
functional $\EuScript{F}$ as $\chi_n^{\ast}\ =\
\frac{\partial\EuScript{F}}{\partial\chi^n}$, where $\chi^n$ is
independent variable. It is straightforward to check that this is
indeed a lagrangian sub-manifold
$$
\delta\chi_{n}^{\ast}\wedge \delta\chi^{n} \big|_{\EuScript{L}}\ =\
\delta\Big(\frac{\partial\EuScript{F}}{\partial\chi^n}\Big) \wedge
\delta\chi_n\ =\ \frac{\partial^2\EuScript{F}}{\partial
\chi^m\partial\chi^n} \delta\chi^m \wedge \delta\chi^n\ =\ 0
$$
since symmetry of the second derivative w.r.t. $m$ and $n$ is
opposite to that of the  $\delta\chi^m \wedge \delta\chi^n$.}
$\EuScript{L}$. The choice of $\EuScript{L}$ generalizes
{\cite{BV,AS}} the standard gauge fixing procedure. As in common
quantum field theory, define effective action integrating out part
of fields. We decompose the space of fields $\chi^n$ in the action
into direct sum of two subspaces. The first one corresponds to
cohomologies of $Q$-operator (the first term in (\ref{Qtot})\ ),
while the second subspace is the orthogonal complement to the first
one. Hence, arbitrary field $\chi^n$ can be written as $\chi^n\ =\
B^n\ +\ b^n$ where $B^n$ is a representative of cohomologies and
$b^n$ stand for the rest of the fields (along these fields we do the
integration). The same decomposition for an antifield gives
$\chi_n^\ast\ =\ B_n^\ast\ +\ b_n^\ast$. Thus effective action is
defined as
\begin{equation}\label{BVintegral}
e^{-S^{eff}(B^n,\ B_n^{\ast})}\ =\ \int\limits_{\EuScript{L}} {D}b\
{D}b^{\ast}\ e^{-\frac{1}{\hbar}S(B^n+b^n,B_n^{\ast}+b_n^{\ast})}
\end{equation}
The integration over quantum fields $b$ and $b^{\ast}$ is restricted
to lagrangian sub-manifold $\EuScript{L}$, index $n$ enumerates
different fields. The crucial observation is that if we start from
BV action (\ref{quantum BV equation}), the BV integral
(\ref{BVintegral}) produces again an effective BV  action
$S^{eff}(B,B^{\ast})$. The following manipulations illustrate this
fact.

  Action $S(B^n+b^n,\ B_n^\ast+b_n^\ast)$ satisfies BV equation
\begin{equation}\label{background laplacian}
0\ =\ \bigtriangleup_{B+b}e^{-\frac{1}{\hbar}S(B^n+b^n,\
B_n^\ast+b_n^\ast)}\ =\ \triangle_B e^{-\frac{1}{\hbar}S(B^n+b^n,\
B_n^\ast+b_n^\ast)}\ +\ \triangle_b e^{-\frac{1}{\hbar}S(B^n+b^n,\
B_n^\ast+b_n^\ast)}
\end{equation}
Since $B$ and $b$ belong to the different spaces, the total Laplace
operator on the space of fields is given by $\triangle_{B+b}\ =\
\frac{\delta}{\delta B^n}\frac{\delta}{\delta B_n^\ast}\ +\
\frac{\delta}{\delta b^n}\frac{\delta}{\delta b_n^\ast}\ =\
\triangle_B\ +\ \triangle_b$. Integrating both sides of equation
(\ref{background laplacian}) over  $\EuScript{L}$ one can get
\begin{equation}\label{background action}
0\ =\ \triangle_B\int\limits_{\EuScript{L}} Db Db^\ast
e^{-\frac{1}{\hbar}S(B^n+b^n,\ B_n^\ast+b_n^\ast)}\ +\
\int\limits_{\EuScript{L}}Db Db^\ast \triangle_b
e^{-\frac{1}{\hbar}S(B^n+b^n,\ B_n^\ast+b_n^\ast)}
\end{equation}
Be there no restriction of integration to $\EuScript{L}$ in the
second term, the integrand would be a total derivative, hence the
integral would be vanishing. However the domain of integration is
restricted to $\EuScript{L}$, hence this argument does not work.
Still, as we will show in a moment, the second term in
(\ref{background action}) is equal to zero due to the fact that this
domain is a lagrangian sub-manifold. Hence, effective action
(\ref{BVintegral}), which is the first term in (\ref{background
action}), again satisfies Master Equation. To see that the second
term vanishes we rewrite it  as an integral over the whole space of
$b$ and $b^\ast$, but with the $\delta$-function inserted, which
restricts the domain of integration to $\EuScript{L}$:
$$
\int Db Db^\ast \delta(b_i^\ast - \frac{\partial\EuScript{
F}}{\partial b^i}) \triangle_b e^{-\frac{1}{\hbar}S(B^n+b^n,\
B_n^\ast+b_n^\ast)}
$$
Here $\EuScript{F}$ - is the generating functional of lagrangian
sub-manifold (see footnote \ref{Lagrang subman definition}).
Integrating out the field $b_i^\ast$, one can get
$$
\int\!Db\  \triangle_b e^{-\frac{1}{\hbar}S(B^n+b^n,\
B_n^\ast+b_n^\ast)}
\Bigg|_{b_i^{\ast}=\frac{\partial\EuScript{F}}{\partial
b^i}}\!\!\!\!\!\!\!\!=\ \int Db \left[
\frac{d}{db^i}\left(\frac{\partial}{\partial
b_i^\ast}e^{-\frac{1}{\hbar}S(B^n+b^n,\ \!
B_n^\ast+b_n^\ast)}\right)\ \!\!-\
\frac{\partial^2\EuScript{F}}{\partial b^i\partial
b^k}\frac{\partial}{\partial b_k^\ast}\frac{\partial}{\partial
b_i^\ast}e^{-\frac{1}{\hbar}S(B^n+b^n,\ \!B_n^\ast+b_n^\ast)}\right]
$$
Here we extracted the total derivative, which upon integration over
$Db$ vanishes. The last expression should be evaluated at
$b_i^\ast=\frac{\partial\EuScript{F}}{\partial b^i}$. Since the
generating functional $\EuScript{F}$ depends only on the fields
$b^i$ and does not depend on antifields $b_i^\ast$ one can twice
integrate by parts the second term in the r.h.s. The result is
$-\int\  Db\ \EuScript{F}\ \frac{\partial^4}{\partial b^i\ \!
\!\partial b^k\ \!\!\partial b_i^\ast\ \!\!
\partial b_k^\ast}\ e^{-\frac{1}{\hbar} S}$. This term is equal to
zero because either $b$ or $b^\ast$ are odd fields.

 This remark concludes the proof that effective action for the BV
action satisfying (\ref{quantum BV equation}), satisfies
(\ref{quantum BV equation}) again. The same statement can be proved
for the classical BV equation (\ref{classical BV equation}) in case
initial action does not depend on $\hbar$. This can be done by
repeating the above arguments in the limit $\hbar \rightarrow 0$.

  In the present paper we do not discuss the general solution of BV
Master  Equation. For our purposes it is enough to know BV action
for two particular examples. They are: BV action for gauge theories,
linear in anti-fields \cite{BV} and AKSZ (Alexandrov, Kontsevich,
Schwarz, Zaboronsky)-type  BV actions \cite{AKSZ}.

{\bf\underline{BV for gauge theories}}\\
Consider a gauge invariant action $I[A]$. The standard way to make
perturbative calculations with this action is to fix the gauge and
introduce Faddeev-Popov ghosts. For the following calculations we
use $R_\xi$ gauge. Thus the action is
\begin{equation}\label{BRST action}
S^{BRST}\ =\ I[A]\ +\ \int B^{a}\partial_{\mu}A_{\mu}^{a} \ -\
\frac{\xi}{2}B^{a}B^{a}\ +\
\overline{c}^{a}(-\partial_{\mu}D_{\mu}^{a c} c^{c})
\end{equation}
where $D_\mu^{a c}\ =\ \partial_\mu\delta^{a c}\ +\ gf^{a b c
}A_\mu^b$ is the covariant derivative\footnote{$a,b,c$ - are color
indices and $f^{a b c}$ - structure constants for algebra
$\mathfrak{g}$.}. We refer to this action as BRST action, because it
is invariant under the symmetry \cite{BRST}
$$
\chi^n\ \rightarrow\ \chi^n \ +\ \epsilon s \chi^n
$$
where $\chi^n$ is an arbitrary field in the theory, $s$ is the
operator $ s\ =\ \int D_{\mu}^{a c}c^c \frac{\delta}{\delta
A_{\mu}^a}\ -\ \frac{1}{2}g f^{a}_{\ b c} c^b c^c
\frac{\delta}{\delta c^{a}}\ +\ B^{a}\frac{\delta}{\delta
\overline{c}^a} $ and $\epsilon$ is infinitesimal odd parameter.  BV
action for this model is
\begin{equation}\label{BVgauge}
S^{BV}\ =\ I[A]\ +\ s\int \chi^n \chi_n^{\ast}\ \ \ =\ \ \ I[A]\ +\
\int D_{\mu}^{a c} c^c (A_{\mu}^a)^{\ast}\ -\ \frac{1}{2}g f^{a}_{\
b c} c^b c^c (c^{a})^\ast\ +\ B^a (\overline{c}^a)^\ast
\end{equation}
It is straightforward to check that this action satisfies  classical
BV equation (\ref{classical BV equation}) under the condition that
$I[A]$ is gauge invariant  action and structure constants $f^{a}_{\
b c}$ for the gauge algebra satisfy Jacobi identity.

  To obtain action (\ref{BRST action}) from (\ref{BVgauge}) one
should make the substitution
\begin{equation}\label{From BV to BRST }
\begin{array}{ccc}
\chi_n^\ast\ =\ \frac{\partial \EuScript{F}}{\partial \chi^n} &\ \ \
$where$\ \ \  & \EuScript{F}\ =\ \int \overline{c}^a
\left(\partial_\mu A_\mu^a\ -\ \frac{\xi}{2}B^a \right)
\end{array}
\end{equation}
for each antifield in (\ref{BVgauge}). This substitution is
equivalent to the choice of lagrangian sub-manifold $\EuScript{L}$
and $\EuScript{F}$-is the generating functional of $\EuScript{L}$
(see footnote \ref{Lagrang subman definition}). Thus, standard
diagram technique in gauge theories, which appears from the integral
of the action (\ref{BRST action}), can be viewed as BV integral
(\ref{BVintegral}) over the lagrangian sub-manifold (\ref{From BV to
BRST }) for the action (\ref{BVgauge}). This simple calculation
illustrates that for particular example of gauge invariant action BV
integration over lagrangian sub-manifold is equivalent to the common
gauge fixing procedure.

 In the literature on BV quantization the action (\ref{BVgauge}) is
usually called BV action with Lagrange multiplier, which is the last
term $B^a (\overline{c}^a)^\ast$. Obviously, BV action without
Lagrange multiplier  is defined as
\begin{equation}\label{BV action without L.m.}
S^{BV}_{\begin{array}{c}$\footnotesize{without
L.m.}$\end{array}}\!\!\!= \ I[A]\ +\ \int D_{\mu}^{a c} c^c
(A_{\mu}^a)^{\ast}\ -\ \frac{1}{2}g f^{a}_{\ b c} c^b c^c
(c^{a})^\ast\
\end{equation}
Such action again satisfies classical BV equation. It turns out that
BV integral with the action (\ref{BV action without L.m.})
restricted to the certain lagrangian sub-manifold is equivalent to
the BV integral with the action (\ref{BVgauge}) restricted to
lagrangian sub-manifold (\ref{From BV to BRST }) in the limit
$\xi\rightarrow 0$.

  For our calculation the theory of Chern-Simons plays important
role. It should be mentioned that in case of Chern-Simons the same
result for BV action (\ref{BV action without L.m.}) without Lagrange
multiplier can be obtained by the following procedure. Suppose that
the field $\Psi$ can be expanded as $\Psi\ =\ \Psi^{(0)}\ +\
\Psi^{(1)}\ +\ \Psi^{(2)}\ +\ \Psi^{(3)}$. Here superscripts
${(0)..(3)}$ stand for the degree of the form. BV action for $3D$
Chern-Simons can be obtained by such expansion from
\begin{equation}\label{CSAKSZ}
S^{BV-CS}\ =\ \int\limits_{X^3} STr \left(\ \Psi d \Psi \ +\
\frac{2}{3}g \Psi^{3} \right)
\end{equation}
under the identification
$$
\begin{array}{cc}
\Psi^{(0)}\ =\ c\ \ \ \ \ \ \ \  & \Psi^{(2)}\ =\ \frac{1}{4}\
\varepsilon^{\mu \nu \lambda}\ A^\ast_\mu\
 dx^\nu\wedge dx^\lambda \\ &   \\
\Psi^{(1)}\ =\ A_\mu\ dx^\mu\ \ \ \ \ \ \ \  &   \Psi^{(3)}\ =\
\frac{1}{24}\ \varepsilon^{\mu \nu \lambda}\   c^{\ast} \
dx^\mu\wedge dx^\nu\wedge dx^\lambda
\end{array}
$$
Here $c$ and $A$ are ghost field and gauge connection and $A^{\ast}$
and $c^{\ast}$ are BV anti-gauge field and BV anti-ghost. It is
straightforward to check that action (\ref{CSAKSZ}) satisfies
classical BV equation (\ref{classical BV equation}) and reproduces
action (\ref{BV action without L.m.}) in case the gauge invariant
action $I[A]\ =\ \int Tr \left(\ A d A \ +\ \frac{2}{3}g A^{3}
\right)$ is $3D$ Chern-Simons action\footnote{\label{supertrace}It
should be mentioned that symbol STr in (\ref{CSAKSZ}) stands for the
supertrace, which under cyclic permutation can change the sign if
odd field passes through it: $STr (A B C) \ = \ (-1)^C STr (C A B)\
=\ (-1)^{C+B} STr (B C A)$.}. This technique (expansion of $\Psi$ in
(\ref{CSAKSZ})) for solving the classical Master Equation can be
generalized from the integration over 3-dimensional manifold to
arbitrary odd dimension.

{\bf\underline{BV for AKSZ}}\\
Another large class of BV actions can be found by procedure similar
to the one mentioned above. Suppose we have some nilpotent operator
$d$ ( though we denote this operator $d$ as de Rham operator, all
results do not depend on its particular form ) and function
$f(\widetilde{\varphi}, \varphi)$, satisfying
\begin{equation}\label{condition on function f}
\frac{\partial f}{\partial\widetilde{\varphi}_i}\frac{\partial
f}{\partial \varphi^i}\ =\ 0
\end{equation}
It can be proved \cite{AKSZ} that under these conditions the action
\begin{equation}\label{AKSZ}
S^{AKSZ}\ =\ \int\limits_{X^d} STr\Big( \Psi^\ast d  \Psi\ +\ \
f(\Psi^\ast, \Psi) \Big)
\end{equation}
is a BV action, when $\Psi\ =\ \Psi^{(0)}\ +\ \Psi^{(1)}\ +\ ...\
+\Psi^{(d)}$ and $\Psi^\ast\ =\ \Psi^{\ast (0)}\ +\ \Psi^{\ast(1)}\
+\ ...\ +\Psi^{\ast(d)}$. In a moment we will illustrate this result
but first we should divide the whole space of fields $\Psi$ and
$\Psi^\ast$ into the pairs: field-antifield. Such pairs are
organized as follows: $\big( \Psi^{(0)}\ ,\ \Psi^{\ast(d)} \big)$,\
$\big( \Psi^{(1)}\ ,\ \Psi^{\ast(d-1)} \big)$,\  $\big( \Psi^{(2)}\
,\ \Psi^{\ast(d-2)} \big)$ and so on. The first element in each pair
is a field, the second one is an antifield. The easiest way to check
that (\ref{AKSZ}) is a BV action is to write it  in the  matrix
form:
$$
S^{AKSZ}\ =\ \Psi^\ast_A d^A_{\ \ B} \Psi^B\ +\ f(\Psi^\ast,\ \Psi)
$$
Then classical BV equation (\ref{classical BV equation}) gives
\begin{equation}\label{Leibnitz1}
\begin{split}
\frac{\delta_L S}{\delta\Psi^B}\frac{\delta_R S}{\delta\Psi^\ast_B}\
=\ (-1)^{C_1}\Psi^\ast_A\ d^A_{\ \ B}\ d^B_{\ \  C}\ \Psi^C\ \ +\ \
(-1)^{C_2}\Psi^\ast_A\ d^A_{\ \ B}\ \frac{\delta_R
f}{\delta\Psi^\ast_B}\ \ +\ \ (-1)^{C_3}\frac{\delta_L
f}{\delta\Psi^B}d^B_{\ \ C}\Psi^C\ \ +\ \ \frac{\delta_L
f}{\delta\Psi^B}\frac{\delta_R f}{\delta\Psi^\ast_B}
\end{split}
\end{equation}
Here $C_1$, $C_2$, $C_3$ are constants which depend on the parities
of the fields (one should consider $d^A_{\ \ B}$ as an odd
operator). These coefficients can be directly determined however
they are not important for our calculations.  In this expression the
first term vanishes due to nilpotency of $d^A_{\ \ B}$, the last
term - due to (\ref{condition on function f}). While the sum of the
second and the third terms is equal to zero due to Leibnitz rule
(see explicit check (\ref{Leibnitz2}) below).

  For particular example of functions  $f(\widetilde{\varphi},\varphi)\ =\ g f^{a}_{\ b
c}\ \widetilde{\varphi}_a\varphi^b\varphi^c$, the action
(\ref{AKSZ}) reproduces the action for AKSZ-Chern-Simons theory
\begin{equation}\label{PCS}
S^{AKSZ-CS}\ =\ \int\limits_{X^d}\ STr\Big(\Psi^\ast d \Psi\ +\ g
\Psi^\ast\Psi^2\Big)
\end{equation}
Condition (\ref{condition on function f}) is satisfied for such
function  $f(\widetilde{\varphi},\varphi)$ because of Jacobi
identity for the structure constants. The sum of the second and
third terms of (\ref{Leibnitz1}) in this particular example gives
\begin{equation}\label{Leibnitz2}
(-1)^{C_4}\ \Psi^\ast_A\  d^A_{\ B}\ f^B_{L M}\  \Psi^L \Psi^M\ \ +\
\ (-1)^{C_5}\ f^K_{B M} \Psi^\ast_K\ d^B_{\ C} \Psi^C\Psi^M\ \ +\ \
(-1)^{C_6}\ f^K_{L B}\ \Psi^\ast_K \Psi^L d^B_{\ C} \Psi^C
\end{equation}
The constants $C_4$, $C_5$, $C_6$ are organized in such a way that
this expression is equivalent to Leibnitz identity ($d(f\cdot g)\ -\
d f\cdot g\ -\ f \cdot dg\ =\ 0$\ ) for the function $f^K_{L
M}\Psi^L\Psi^M$. Hence, if operator $d^A_{\ B}$ satisfies Leibnitz
identity expression (\ref{Leibnitz2}) is equal to zero.

\newpage

\noindent{\bf Why are we interested in such a strange object as AKSZ
action?}
\begin{enumerate}
\item The first reason is that our fundamental theory (\ref{Fund})
is exactly AKSZ theory. This is true since $Q_B$ is nilpotent
operator (similarly to operator $d$ in (\ref{AKSZ})) in the space
$O[\theta_\alpha\ \! ,\ \! \lambda_\alpha\ |\
f^{\mu}(\lambda)]\otimes Func(x)$. This operator also satisfies
Leibnitz identity because it is linear differential operator of
first order in derivatives (
$\frac{\partial}{\partial\theta_\alpha}$ and $\partial_\mu$). The
field $\EuScript{A}$ in (\ref{Fund}) should be identified with
$\Psi$ and the field $\EuScript{P}$ with AKSZ antifield $\Psi^\ast$.

\item The second reason is that for {\bf "regular"} system of
quadrics effective action for the theory (\ref{Fund}) is precisely
AKSZ version of Chern-Simons theory. This fact will be proved in
chapter 5.

\end{enumerate}

 It should be mentioned that in odd space-time dimensions there is
$Z_2$ symmetry between the fields $\Psi$ and $\Psi^\ast$ in the
action which preserves the BV structure. To see this take as an
example $d=1$. The expansion of fields $\Psi$ and $\Psi^\ast$ in
this case gives $\Psi\ =\ \Psi^{(0)}\ +\ \Psi^{(1)}\ =\ c\ +\ A\
\!\! dx$ and $\Psi^\ast\ =\ \Psi^{\ast(0)}\ +\ \Psi^{\ast(1)}\ =\
A^\ast\ +\ c^{\ast}\ \!\! dx$, where $c$, $A$, $A^\ast$, $c^\ast$
are BV fields while $dx$ is the generator of exterior algebra in
$d=1$. Since the field $\Psi$ has certain total parity (the parity
of a BV field plus the parity of $dx$ which is $1$), the fields $c$
and $A$ have opposite internal BV parities. On the other hand, the
field $c$ has opposite parity to $c^\ast$, which is its antifield.
From these two facts we conclude that the fields $\Psi$ and
$\Psi^\ast$ have equal total parities. This is true in odd
space-time dimensions. Hence, there is the symmetry
$$
\begin{array}{c}
\Psi\ \longrightarrow\  \Psi^\ast\\
\Psi^\ast\ \longrightarrow\  \Psi
\end{array}
$$
of AKSZ action (\ref{PCS}) and BV structure. This is $Z_2$ symmetry
of the construction. In case of even space-time dimensions there is
no such symmetry because the fields $\Psi$ and $\Psi^\ast$ have
opposite parities.  Factorizing over this $Z_2$ symmetry  in odd
dimension one can obtain\footnote{More precisely this is true under
the change $g\rightarrow\frac{2}{3}g$.} the gauge theory of
Chern-Simons (\ref{CSAKSZ}) from the AKSZ version of Chern-Simons
(\ref{PCS}). This reduction $\Psi^\ast\ =\ \Psi$ kills one half of
degrees of freedom in AKSZ-Chern-Simons theory.

\begin{figure}[h]
\centerline{\includegraphics[width=64mm]{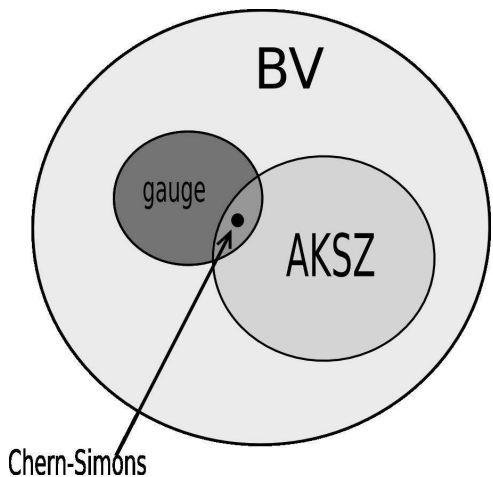}}
\caption{{\footnotesize A primitive illustration on the structure of
BV space. Chern-Simons theory can be obtained both from the BV
formalism for gauge theories and from the reduction of AKSZ version
of Chern-Simons. }} \label{fig1}
\end{figure}

The  main ideas of this section are presented in the figure
\ref{fig1}.
\vspace{-0.2cm}\noindent The largest circle is the whole space of BV
theories which we do not discuss in the present paper. One subclass
of this space is BV for gauge theories, another one - AKSZ-type
theories. In the intersection of these two subclasses the BV version
of Chern-Simons is located.

 The main issue of this section is to introduce some terminology,
which is probably not common, but is widely used  for the following
discussion. In the next chapters we will demonstrate that a large
class of AKSZ theories and even gauge theories  can be obtained as
effective BV actions from $S^{Fund}$ of (\ref{Fund}). \vspace
{-0.2cm}

\section{Quantum calculations.}\label{Quantum calculations}
 In this chapter we explain how one should conduct Feynman
diagram calculations in BV integral (\ref{BVintegral}) starting from
the action $S^{Fund}$ of (\ref{Fund}). Since operator $Q_{B}$ is
nilpotent, it is clear that the action (\ref{Fund}) has a large
gauge freedom. This freedom is to be fixed by the choice of certain
lagrangian sub-manifold in the integral (\ref{BVintegral}) providing
a well defined diagrammatic expansion for $S^{eff}$. Since initial
action (\ref{Fund}) contains Grassmann variables in the definition
of fields, the diagram expansion of integral (\ref{BVintegral})
terminates at some point, leaving only finite number of diagrams.
These diagrams can be summed up explicitly into a very simple
effective action, which under certain conditions represents some
interesting theories.

 The fields in the action (\ref{Fund}) belong to the spaces
$O[\theta_\alpha\ \! ,\ \! \lambda_\alpha\ |\
f^{\mu}(\lambda)]\otimes Func(x)\otimes T(\mathfrak{g})$ and
$O[\theta_\alpha\ \! ,\ \! \lambda_\alpha\ |\
f^{\mu}(\lambda)]^\ast\otimes Func(x)\otimes T(\mathfrak{g})$. For
the discussion in  this section we omit factors $Func(x)$ and
$T(\mathfrak{g})$, since they are inessential. The only relevant
information is how the fields depend on $\lambda$ and $\theta$.

 Expand the fields $\EuScript{A}$ and $\EuScript{P}$ in the action
(\ref{Fund}) in the classical and quantum parts $\EuScript{A}\ =\
\mathsf{A}\ +\ a$ and $\EuScript{P}\ =\ \mathsf{P}\ +\ p$, where
classical fields $\mathsf{A}$ and $\mathsf{P}$ are chosen to take
value in the space of cohomologies $\mathcal{H}(Q)$ of the operator
$Q=\lambda_{\alpha}\frac{\partial}{\partial\theta_{\alpha}}$ which
is the first term in (\ref{Qtot}), and  quantum fields $a$ and $p$
are from the orthogonal complement to $\mathcal{H}(Q)$. Along these
fields we do the integration. Then, the action (\ref{Fund}) can be
rewritten as
\begin{equation}\label{Sdiagr}
\begin{split}
S^{Fund}(\mathsf{P}+p,\mathsf{A}+a)\ =\ \int\ STr \bigg( <p,Qa>\ +\
<\mathsf{P},\Phi a>\ +\ <\mathsf{P},\Phi \mathsf{A}>\ +\ <p,\Phi
\mathsf{A}>\ +\\ +\  <p,\Phi a>\ +\
g<(\mathsf{P}+p),(\mathsf{A}+a)^2> \bigg)
\end{split}
\end{equation}
Since external fields $\mathsf{A}$ and $\mathsf{P}$ belong to the
representatives of $\mathcal{H}(Q)$, the terms $<\mathsf{P},\
Q\mathsf{A}>$, $<\mathsf{P},\  Q a>$, $<p,\  Q\mathsf{A}>$ are equal
to zero. It is straightforward to extract vertices from this action.
They are:
\begin{figure}[h]
\centerline{\includegraphics[width=157mm]{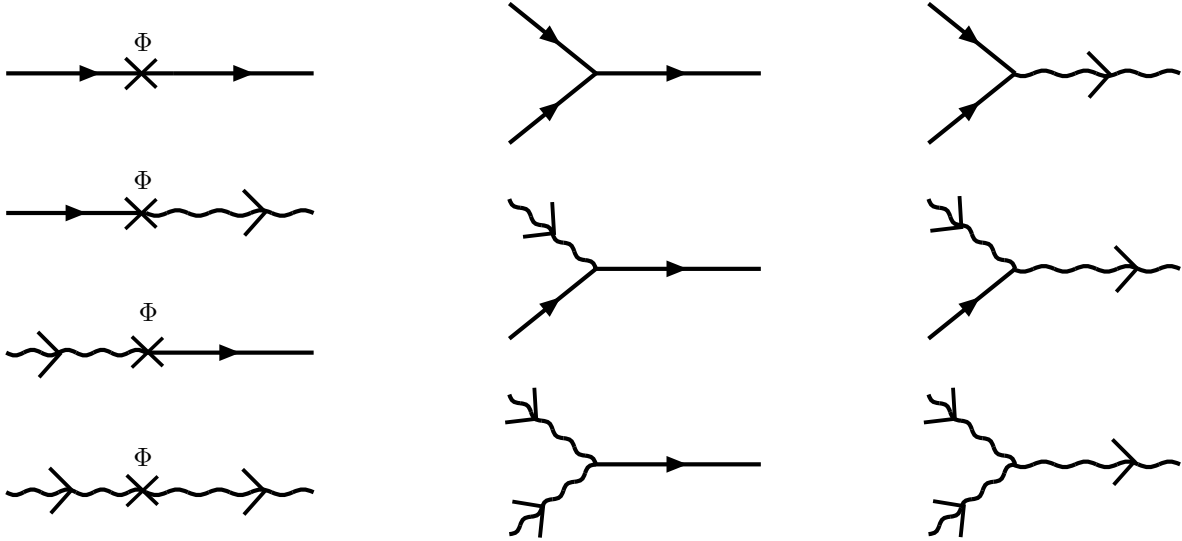}}
\caption{{\footnotesize All the vertices in the theory
(\ref{Sdiagr}).}} \label{vertices}
\begin{picture}(600,0)(-20,140)
\put(48,203){$\Phi$}
\put(50,258){$\Phi$}\put(48,308){$\Phi$}\put(48,360){$\Phi$}
\end{picture}
\end{figure}

In these diagrams solid lines represent external fields. Wavy lines
stand for the propagating quantum field. Each line is provided with
the arrow sign which notifies whether this line belongs to a field
($\EuScript{A}$) or an antifield ($\EuScript{P}$) of BV formalism.
For each field the arrow points into the diagram, for each antifield
the arrow points out of the diagram. Each 3-valent vertex
contributes a factor of $g$ - the coupling constant.

  The  first term in (\ref{Sdiagr}) is  responsible for the
propagator of the quantum field. Since operator $Q$ is nilpotent it
has a large number of zero modes. To eliminate them consider the
following construction\footnote{As we explained we chapter 2 the
correct procedure for evaluation of the propagator is the choice of
lagrangian sub-manifold. Below we show that such choice kills all
the zero modes in the kinetic term.}. Define the space $V$ as a
complement of $\mathcal{H}(Q)$ in $O[\theta_\alpha\ \! ,\ \!
\lambda_\alpha\ |\ f^{\mu}(\lambda)]$
$$
O[\theta_\alpha\ \! ,\ \! \lambda_\alpha\ |\ f^{\mu}(\lambda)] =\
\mathcal{H}(Q)\ \oplus\ V
$$
There is a subspace $\mathcal{C}$ in $V$ of all closed forms which
are exact ($\mathcal{C}\subset V$). Choose a basis elements
$c_\alpha\in\mathcal{C}$, then
\begin{equation}\label{ceal}
c_{\alpha}\ =\ Q(b_{\alpha})
\end{equation}
In this expression, the elements $b_{\alpha}$ are defined modulo
$Q$-exact expressions, remaining the ambiguity in the choice of
$b_\alpha$. Define the operator $K$ as
\begin{equation}\label{K}
{\begin{array}{c}
K c_{\alpha}\ =\ b_{\alpha}\\
K b_{\alpha}\ =\ 0
\end{array}}
\end{equation}
This operator maps each element $c_{\alpha}$ to certain element
$b_{\alpha}$ breaking the ambiguity of (\ref{ceal}). From
(\ref{ceal}) and (\ref{K}) it is obvious that $\{K,\ Q\}\ =\ 1$ on
the space spanned by $c_{\alpha}$ and $b_{\alpha}$. It is
straightforward to check that this space coincides with $V$. Indeed,
for arbitrary element $v \in V$, the element $Qv$ is $Q$-closed.
Hence, $Qv\ =\ x_{\alpha}c_\alpha$ with some coefficients
$x_\alpha$. Thus, the element $v$ can be represented as
\begin{equation}\label{basis}
v\ =\ x_{\alpha}b_{\alpha}\ +\ y_{\alpha}c_{\alpha}
\end{equation}
where  $y_{\alpha}$ are some coefficients. \\

  Conducting evaluation of the integral (\ref{BVintegral}) with the action
(\ref{Sdiagr}) we have already removed part of zero modes from the
kinetic term, absorbing them into external fields $\mathsf{P}$ and
$\mathsf{A}$. The point is that some of zero modes still survived,
because the integral is over the space $V$ which has a sub-space
$\mathcal{C}\subset V$ of closed forms. Thus to determine the
propagator of  quantum field we should remove this ambiguity also.
From (\ref{basis}) it is clear that demanding $Kv\ =\ 0$ kills all
the terms proportional to $c_{\alpha}$, removing the remnant of zero
modes. Thus, the integration in (\ref{BVintegral}) should be
restricted by the condition\footnote{Since $p$ is the element of the
dual space (see footnote \ref{dual}) operators $Q$ and $K$ act on it
from the right.}
\begin{equation}\label{Lagr}
{\begin{array}{c}K a\ =\ 0\\
p K\ =\ 0
\end{array}}
\end{equation}
The only question is whether this condition is a choice of
lagrangian sub-manifold or not. The answer to this question is
positive. Since $\{K,\ Q\}\ =\ 1$ in the space $V$, both $Q$ and $K$
have no cohomologies in this space. Hence $Ka\ =\ 0$ implies $a\ =\
Kw$ for some element $w$ and
\begin{equation}
\int  Tr <\delta p\ ,\  \delta a>\ =\ \int Tr <\delta p\ ,\ K \delta
w>\ =\ \int Tr <\delta(p K)\ ,\  \delta w>\ =\ 0
\end{equation}
which is the definition of lagrangian sub-manifold (see footnote
\ref{Lagrang subman definition}).

  Now the integral (\ref{BVintegral}) can be rewritten as
\begin{equation}\label{BVVintegral}
e^{-S^{eff}(\mathsf{P}, \mathsf{A})}\ =\
\int\limits_{\begin{array}{c} Ka=0\\ pK = 0
\end{array}}
\!\!\!{D}p {D}a\ e^{-S^{Fund}(\mathsf{P}+p,\  \mathsf{A}+a)}
\end{equation}
This is BV integral, providing $S^{Fund}$ is BV action and
(\ref{Lagr}) determines a lagrangian sub-manifold. Hence
$S^{eff}(\mathsf{P}, \mathsf{A})$ is again BV action for some
theory. Explicit expression for $S^{eff}(\mathsf{P}, \mathsf{A})$ is
encoded in the choice of background fields $\mathsf{P,A}$, hence in
cohomologies of operator $Q$. These cohomologies for an arbitrary
set of constraints $f^{\mu}(\lambda)$ will be calculated in the next
section.

\section{Calculation of Cohomologies}\label{Cohomology evaluation}
  In this chapter we present the general procedure, which allows to find all
cohomologies of $Q$-operator in the space $O[\theta_\alpha\ \! ,\ \!
\lambda_\alpha\ |\ f^{\mu}(\lambda)]$. This result, to the best of
our knowledge is new. Subsection \ref{idea} contains the general
idea, subsection \ref{proof} - the brute force proof of the theorem
(\ref{coh}).
\subsection{Procedure}\label{idea}
  The space of complex
$O[\theta_\alpha\ \! ,\ \! \lambda_\alpha\ |\ f^{\mu}(\lambda)]$ is
defined by the set of quadratic constraints $f^{\mu}(\lambda)$. The
function $G^{(1)\ \mu}(\lambda)$ is called a {\bf relation} if
\begin{equation}\label{relation}
\sum\limits_{\mu=1}^{N}\ \  G^{(1)\ \mu}(\lambda)\ f_{\mu}(\lambda)\
=\ 0
\end{equation}
The meaning of superscript $(1)$ will be defined below. From the
definition it is clear that multiplying relation $G^{(1)\ \mu}$ by
an arbitrary  element of the ring $C[\lambda]$ of polynomials in
$\lambda_\alpha$, we obtain again a relation. Hence, the space of
functions $G^{(1)\ \mu}$, satisfying equation (\ref{relation}) forms
a module over the ring $C[\lambda]$. The minimal set of functions
$G_A^{(1)\ \mu}(\lambda)$ (index $A$ enumerates different functions)
which generates the whole module via multiplication by the elements
of $C[\lambda]$ is called the {\bf fundamental system of relations}.
An arbitrary relation $G^{(1)\ \mu}(\lambda)$ can be obtained from
the fundamental system of relations $G_A^{(1)\ \mu}(\lambda)$ as the
sum $G^{(1)\ \mu}(\lambda)\ =\ K^A(\lambda)G_A^{(1)\ \mu}(\lambda)$,
where $K^A(\lambda)$ is some function. In case $K^A(\lambda)\ =\
const$, relation $G^{(1)\ \mu}(\lambda)$ is fundamental. If
$K^A(\lambda)\ \neq\ const$ relation $G^{(1)\ \mu}(\lambda)$ is not
fundamental.

\noindent Relation $G^{(1)\ \mu}(\lambda)$ is called {\bf trivial}
if it belongs to the ideal $I_f$ ( if its coefficients are
proportional to
$f^{\mu}(\lambda)$ $ $).  \\

\noindent The following  example illustrates the definitions given
above.

\noindent{\bf\underline{Example}} Consider the set of quadrics
$f_1=\lambda_1^2\ \ \ f_2=\lambda_1\lambda_2$. Relation
$G^\mu(\lambda)\ \ $ ($G^{\ 1}f_1+G^{\ 2}f_2=0$) is trivial for
$$
G^{\ 1}=\lambda_1\lambda_2\ \ \ \ G^{\ 2}=-\lambda_1^2
$$
because its coefficients are proportional to quadrics.

\noindent Relation $L^\mu(\lambda)$:
$$
L^{\ 1}=\lambda_2\ \ \ \ \ L^{\ 2}=-\lambda_1
$$
is non-trivial.

\noindent In this particular example relation $L^\mu(\lambda)$ is
fundamental, while $G^\mu(\lambda)$ is not fundamental (it is
obtained from multiplication of $L^\mu(\lambda)$ by $\lambda_1$). In
general situation this is not true. Whether relation is trivial or
not has nothing to do with the fact whether it is fundamental or
not. The fundamental system of relations for this set of quadrics
$f^\mu(\lambda)$ consists of one relation $G_1^{(1)\ \mu}$ which
coincides with $L^\mu(\lambda)$\ \
$(\ G_1^{(1)\ \mu}\ =\ L^\mu(\lambda)\ )$. \\

  Suppose  the fundamental system of relations is found. It is
given by the set of basis elements enumerated by the index $A$.
Define the {\bf second-level relations}  as "relations for
relations":
\begin{equation}\label{secondary relation}
\sum\limits_{A}\ \  G^{(2)\ A}(\lambda)\ G_{A}^{(1)\ \mu}(\lambda)\
=\ 0
\end{equation}
Now the meaning of the superscripts $(1)$ and $(2)$ is obvious.
Equation (\ref{secondary relation}) should be valid for arbitrary
value of index $\mu$. Such secondary relations again form a module.
It is possible to find a basis in this module - the fundamental
system of relations at the second level. These relations are
$G_B^{(2)\ A}(\lambda)$, where index $B$ enumerates the fundamental
relations at the second level. Conducting the same manipulations, we
define $G_C^{(n)\ D}(\lambda)$ at the n-th level. The
following theorem can be proved:\\
{\bf\underline{Theorem}} Suppose the fundamental set $G_A^{(n)\
B}(\lambda)$ is found. All cohomologies $\mathcal{H}(Q\ ,\ O)$ of
the $Q$-operator in the space $O[\theta_\alpha\ \! ,\ \!
\lambda_\alpha\ |\ f^{\mu}(\lambda)]$ are given by:
\begin{eqnarray}\label{coh}
& 1 &   \nonumber\\
&  \widetilde{D} f^{\mu}(\lambda)  & \nonumber\\
& \widetilde{D}\Big( \widetilde{D}G_{A_{1}}^{(1)\ \mu}(\lambda)\  f_{\mu}(\lambda)\Big)& \\
& \widetilde{D} \Bigg(\widetilde{D} \Big(\widetilde{D} G_{A_{2}}^{(2)\ A_{1}}(\lambda)\ G_{A_{1}}^{(1)\ \mu}(\lambda)\Big)\  f_{\mu}(\lambda)\Bigg)& \nonumber \\
&...........................................................&
\nonumber
\end{eqnarray}
We assume summation over repeated indices. To define operator
$\widetilde{D}$ consider first nilpotent operator $D$:
$$
D=\theta_a\frac{\partial}{\partial\lambda_a}
$$
It is straightforward to check that anticommutator
\begin{equation}\label{deg-old}
\{Q\ ,\ D\}\ =\ \theta_a\frac{\partial}{\partial\theta_a}\ +\
\lambda_a\frac{\partial}{\partial\lambda_a}\ =\
deg(\theta)+deg(\lambda)\ =\ deg
\end{equation}
is the degree operator. Define operator $\widetilde{D}$ as
$\widetilde{D}\ =\ \frac{1}{deg}D$. Such operator can be applied to
all functions except constants. In the tower (\ref{coh}) it is
applied to functions $f^\mu(\lambda)$, which are quadratic in
$\lambda_\alpha$ and to the basis functions of fundamental system of
relations. Such functions should be at least linear in
$\lambda_\alpha$. (We are not interested in constant relations
because they are simple linear redefinitions of basis elements. Such
redefinitions do not affect the cohomologies.) Using this definition
equation (\ref{deg-old}) can be rewritten  as
\begin{equation}\label{deg}
\{Q\ ,\ \widetilde{D}\}\ =\ 1
\end{equation}

  The cohomologies (\ref{coh}) are written in such a way, that each new line
gives one extra degree of $\theta$. Since $\theta$ are Grassmann
variables and the number of them is limited in the space
$O[\theta_\alpha\ \! ,\ \! \lambda_\alpha\ |\ f^{\mu}(\lambda)]$,
the tower (\ref{coh}) stops at some point. Hence the number of
representatives is finite.

\subsection{Proof of the theorem}\label{proof}
  The proof of the theorem goes in 3 steps: firstly we check that
expressions (\ref{coh}) are closed, secondly that they are not
exact, finally that (\ref{coh}) gives all cohomologies. The first
two steps are presented in this section, while the third one is
given in the appendix. In this section we denote by equality sign
$=$ the equality between the polynomials in the ring
$C[\lambda,\theta]$, while by $\sim$ sign we mean the equality
modulo factorization over the ideal $I_f$.

\vspace{-0.3cm}
\subsubsection{Closeness}
\noindent 1 is obviously $Q$-closed. Applying $Q$ to second
representative $\widetilde{D} f^{\mu}$, one can get:
\begin{equation}\label{Df closed}
Q \widetilde{D} f^{\mu}\ =\ \{Q,\widetilde{D}\} f^{\mu}\ -\
\widetilde{D} Q f^{\mu}\ =\ f^{\mu}\ \sim\ 0
\end{equation}
In the second term of the first equality the fact that $f^{\mu}$
does not depend on $\theta$, hence $Q f^{\mu}=0$ is used. We also
take into account equation (\ref{deg}). In the last equality we use
the fact that we are working in the space $O[\theta_\alpha\ \! ,\ \!
\lambda_\alpha\ |\ f^{\mu}(\lambda)]$, hence all elements from the
ideal $I_f$ (proportional to $f^{\mu}$) are equivalent to $0$.
Applying $Q$ to the next term,
$$
Q\Bigg[\widetilde{D} \Big( \widetilde{D}G_{A}^{(1)\ \mu}\
f_{\mu}\Big)\Bigg]\ =\ \widetilde{D} G_{A}^{(1)\ \mu} f_{\mu}\ -\
\widetilde{D}\Big( Q \widetilde{D}G_A^{(1)\ \mu}\ f_\mu \Big)\ =\
\widetilde{D} G_{A}^{(1)\ \mu} f_{\mu}\ -\
\widetilde{D}\Big(G_{A}^{(1)\ \mu} f_{\mu}\Big)\ =\ \widetilde{D}
G_{A}^{(1)\ \mu} f_{\mu}\ \sim\ 0
$$
In the first equality we use equation (\ref{deg}), Leibniz identity
for $Q$ and the fact that $f^\mu$ do not depend on $\theta$, hence
$Q f^\mu\ =\ 0$. In the second equality we again explore equation
(\ref{deg}) and $Q G_{A}^{(1)\ \mu}\ =\ 0$. In the next equality we
use that $G_{A}^{(1)\ \mu}\ f_{\mu}\ =\ 0$ is a relation.  Last
equivalence sign is the factorization procedure.

The closeness of other terms in (\ref{coh}) can be proved in a
similar way. \vspace{-0.3cm}
\subsubsection{Non-Exactness}
\noindent 1 is obviously not Q exact. Below we present an
illustration of proof that all other terms in (\ref{coh}) are not
exact. The proof is universal for all the terms in the tower
(\ref{coh}). We take the third term to show how it works. Suppose it
is exact. Since we are working in a coset space this means(we omit
index $A_1$)
$$
\widetilde{D}\Big( \widetilde{D}G^{(1)\ \mu}(\lambda)\
f_{\mu}(\lambda)\Big)\ =\ Q(M)\ +\ c^\mu(\lambda, \theta^2)f_\mu
$$
Here $c^\mu(\lambda, \theta^2)$ are some functions quadratic in
$\theta^\alpha$ (this degree is dictated by the degree in
$\theta^\alpha$ of the l.h.s.). Applying operator $Q$ to this
relation, one can get
$$
\widetilde{D}G^{(1)\ \mu} f_\mu\ -\
\widetilde{D}\Big(Q\widetilde{D}G^{(1)\ \mu}\ f_\mu\Big)\ =\
Qc^\mu(\lambda, \theta^2) f_\mu
$$
Using equation (\ref{deg}) one can show that the second term in the
l.h.s. vanishes. Hence, there is a relation on $f^\mu(\lambda)$
$$
\big[\widetilde{D}G^{(1)\ \mu}\ -\ Qc^\mu(\lambda, \theta^2)\Big]
f_\mu\ =\ 0
$$
Each relation can be expanded in the basis of fundamental relations
$$
\widetilde{D}G^{(1)\ \mu}\ -\ Qc^\mu(\lambda, \theta^2)\ =\
K^A(\lambda, \theta)G_{A}^{(1)\ \mu}
$$
were $K^A(\lambda, \theta)$ is linear in $\theta^\alpha$. Applying
to the both sides operator $Q$, one can get
$$
G^{(1)\ \mu}\ =\ Q\Big(K^A(\lambda, \theta)\Big) G_{A}^{(1)\ \mu}
$$
Since $K^A(\lambda, \theta)$ is linear in $\theta^\alpha$, from this
expression it is clear that $G^{(1)\ \mu}$ is at least one degree in
$\lambda^\alpha$ higher than the fundamental system of relations
$G_{A}^{(1)\ \mu}$, hence is not fundamental. This completes the
proof that the third element in (\ref{coh}) is not exact. The proof
for all other terms is analogous.

  Thus, we have checked that all the expressions which are written in
(\ref{coh}) are indeed cohomologies. The only question remains
whether we have found all of them.

\subsubsection{Completeness} The proof that the set (\ref{coh})
is complete is presented in the appendix.
\bigskip

\noindent Concluding this section, we introduce one more definition.\\
{\bf \underline{Definition of "regular" and "singular" quadrics}}.
Suppose the set of quadrics  $f^{\mu}(\lambda)$ is chosen in such a
way, that all relations, used to generate the tower of cohomologies
(\ref{coh}) are trivial (according to the definition above the
example in section 4.1). In this case we call the set
$f^{\mu}(\lambda)$ - the  {\bf "regular"} system of quadrics. In
case there is at least one non-trivial relation, the system
$f^\mu(\lambda)$ is called {\bf "singular"}.

  From the point of the discussed construction {\bf "regular"}
quadrics are the simplest systems. In the next section it will be
demonstrated that in case of {\bf "regular"} system the algebra of
cohomologies $\mathcal{H}(Q)$ is isomorphic to the exterior algebra
$\Lambda^{\bullet}\{\widetilde{D}f_1,.....,\widetilde{D}f_N\}$ of
the generators $\widetilde{D}f_1,.....,\widetilde{D}f_N$. Hence, the
only information which is inherited by this algebra from the
quadrics $f^\mu(\lambda)$ is that the system is {\bf "regular"} and
that there are $N$ quadrics. All the information about particular
structure ( $\lambda_\alpha$-dependence ) of quadrics is irrelevant
for  $\mathcal{H}(Q)$ and hence for the effective action. Thus,
effective action for arbitrary {\bf "regular"} system of quadrics is
universal, i.e. does not depend on the particular, probably very
complicated, expressions  for $f^\mu(\lambda)$. It happens that this
universal effective action is nothing but BV version of
AKSZ-Chern-Simons theory (\ref{PCS}).

\section{AKSZ version of Chern-Simons as effective BV theory}\label{Chern-Simons
chapter}
  We start from one of the simplest examples. Consider the space
$O[\theta_\alpha\ \! ,\ \! \lambda_\alpha\ |\ f^{\mu}(\lambda)]$,
generated by three $\theta$, three $\lambda$ ($\alpha=1,2,3$) and
the set of quadratic constraints \vspace{-0.2cm}
\begin{eqnarray}\label{relations for CS }
f_{1}=\frac{\lambda_1^{2}}{2}\nonumber \\
f_{2}=\frac{\lambda_2^{2}}{2} \\
f_{3}=\frac{\lambda_3^{2}}{2}\nonumber
\end{eqnarray}
\vspace{-0.9cm}
\subsection{Calculation of Cohomologies}
We begin analysis of this example from calculation of all
cohomologies $\mathcal{H}(Q,\ O)$, according to (\ref{coh}). First
of all one should determine the fundamental set of relations
(\ref{relation}) at the first level $G^{(1)\ \!\mu}_{A}(\lambda)$.
These relations are:
\begin{equation}
\begin{array}{ccccccccccc}
\lambda_3^2 f_2  & - & \lambda_2^2f_3 & = & 0 & \ \ \ \ \ \ \  & G^{(1)}_1\ =\  &\!\!\! (\ \ \ \ 0 &\ \  \lambda_3^2 & -\lambda_2^2\ ) &  \\
\lambda_1^2 f_3  & - & \lambda_3^2f_1 & = & 0 & \ \ \ \ \ \ \  & G^{(1)}_2\ =\  & (\ \ -\lambda_3^2 &\ \  0 &\ \  \lambda_1^2\ ) &  \\
\lambda_2^2 f_1  & - & \lambda_1^2f_2 & = & 0 & \ \ \ \ \ \ \  & G^{(1)}_3\ =\ & (\ \ \ \lambda_2^2 &\ \  -\lambda_1^2 &\ \ \  0\ \ ) &  \\
\end{array}
\end{equation}
There is only one second-level relation $G^{(2)\
\!\!A}_{1}(\lambda)$:
\begin{equation}
\lambda_1^2G^{(1)}_1\ +\ \lambda_2^2G^{(1)}_2\ +\
\lambda_3^2G^{(1)}_3\ =\ 0
\end{equation}
written in the basis $(f_1,\ f_2\ f_3)$. At this point the tower of
relations stops. All found relations $G^{(1)}$ and $G^{(2)}$ are
trivial, because their coefficients are proportional to
$f^{\mu}(\lambda)$, hence the set (\ref{relations for CS }) is "{\bf
regular}", according to the definition in the end of the previous
section. Now we can write down all cohomologies $\mathcal{H}(Q,\ O)$
in this example. They are obtained (\ref{coh}) by applying operator
$\widetilde{D}$ to all fundamental relations\footnote{For example,
the representative of $\mathcal{H}(Q)$ generated by
$G_1^{(1)}(\lambda)$ is $\widetilde{D}\lambda_3^2\widetilde{D}f_2\
-\ \widetilde{D}\lambda_2^2\widetilde{D}f_3\ =\
-\lambda_2\lambda_3\theta_2\theta_3$, which is the last term in the
third line of (\ref{cohCS}).}. The result is:
\begin{equation}\label{cohCS}
\mathcal{H}(Q,\ O)\ =\ \left({\begin{array}{ccc}
     &   1    &     \\
\lambda_1\theta_1  &   \lambda_2\theta_2   & \lambda_3\theta_3  \\
 \lambda_1\lambda_2\theta_1\theta_2 &  \lambda_1\lambda_3\theta_1\theta_3\   & \lambda_2\lambda_3\theta_2\theta_3 \\
     & \lambda_1\lambda_2\lambda_3\theta_1\theta_2\theta_3 &     \\
\end{array}}\right)
\end{equation}

\subsection{An Illustrative Example on Diagram Evaluation}
\begin{figure}[h]
\begin{picture}(600,60)(20,270)
\Photon(220,300)(280,300){-1}{4} \ArrowLine(190,275)(221,300)
\ArrowLine(190,325)(221,300) \ArrowLine(280,300)(311, 325)
\ArrowLine(311,275)(280,300)\put(205,275){$\mathsf{A}_{1}$}
\put(205,322){$\mathsf{A}_{2}$}\put(292,275){$\mathsf{A}_{3}$}
\Line(248,306)(255,299) \Line(248,294)(255,299)
\end{picture}
\caption{{\footnotesize 4-point amplitude.}} \label{4-point}
\end{figure}
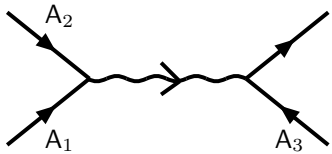

  Second step is evaluation of effective action (\ref{BVVintegral})
by means of diagram expansion in (\ref{Sdiagr}). To begin with we
present a simple example calculating the 4-point amplitude presented
in  figure \ref{4-point}.

As we explain below, this diagram is forbidden (is equal to zero)
for the regular set(\ref{relations for CS }). It is presented here
only for illustrative purposes, clarifying the whole procedure of
diagram evaluation.

\begin{figure}[h]
\centerline{\includegraphics[width=60mm]{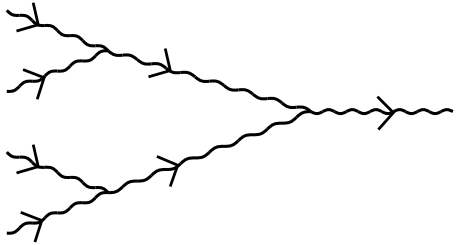}}
\caption{{\footnotesize A tree diagram in the theory
(\ref{Sdiagr}).}} \label{tree}
\end{figure}

 According to discussion after figure \ref{vertices} external lines stand for
the background fields $\mathsf{P,A}\in\mathcal{H}(Q,O)\otimes
Func(x)\otimes T(\mathfrak{g})$. In this particular example there
are three in-lines, corresponding to a  field $\mathsf{A}$, and one
out-line, corresponding to an antifield $\mathsf{P}$.
\noindent It happens that this is the general feature of arbitrary
tree diagram in the theory (\ref{Sdiagr}). Each connected diagram
(only connected diagrams are relevant for calculation of effective
action) can have many in-lines but only one out-line. This is
illustrated in the figure \ref{tree}.

\noindent Due to this fact, the diagram technique is non-trivial
only at the tree level and in 1-loop. All higher loops are absent.
Indeed, the only possibility to organize a loop is to close the
out-line onto one of the in-lines, like in the figure \ref{closing a
tree}.
\begin{figure}[h]
\centerline{\includegraphics[width=50mm]{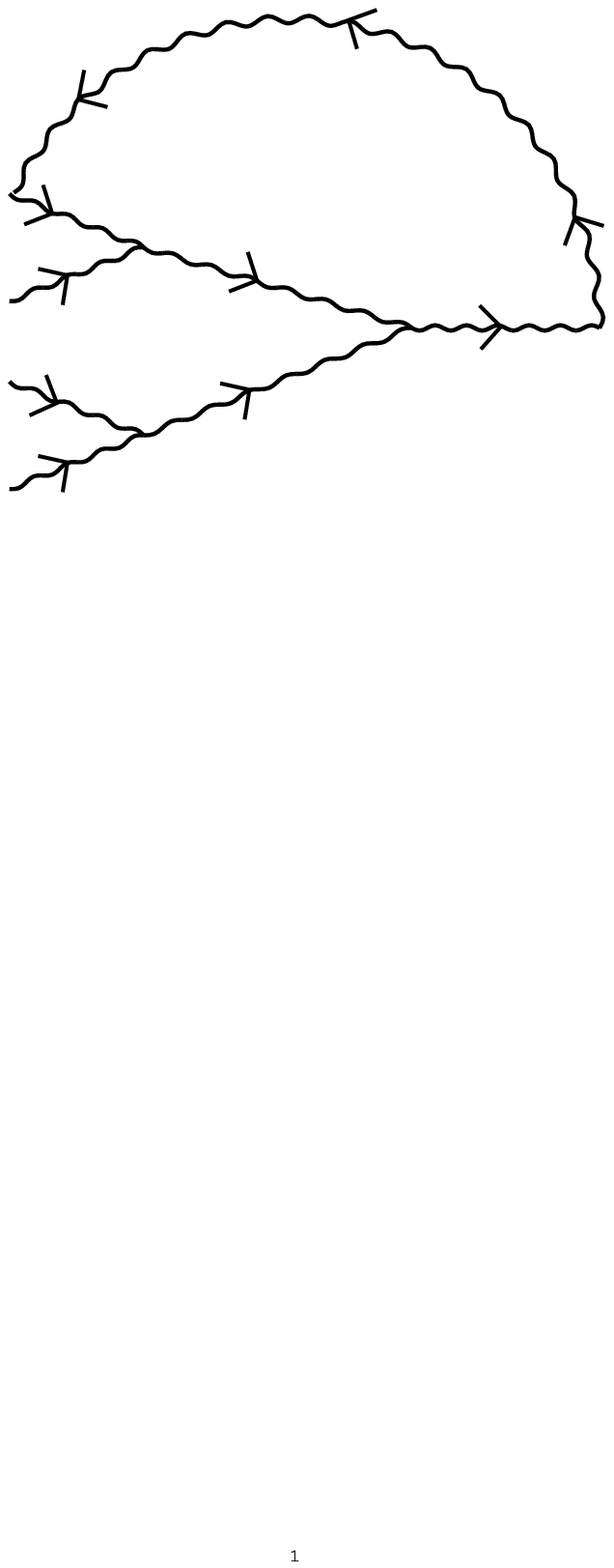}} \caption[
]{\label{closing a tree}  {\footnotesize Closing a tree into one
loop.}}\noindent
\end{figure}
The resulting diagram has only one loop and only in-lines,
prohibiting formation of an another loop.
\newpage

Now we are coming back to evaluation of the diagram in figure
\ref{4-point}. To find this amplitude, we need expression for the
propagator. To obtain it, introduce the sources  for the quantum
fields $a$ and $p$.
\begin{equation}\nonumber
\begin{split}
\int\limits_{\small\begin{array}{c} Ka=0\\ pK = 0\end{array}} DpDa\
e^{-(pQa\ +\ pJ_{p}\ +\ J_{a}a)}\ \ =\ \
\int\limits_{\small\begin{array}{c} Ka=0\\ pK = 0\end{array}} DpDa\
e^{-(pQa\ +\ pJ_p\ +\ J_a \{K,Q\}a)}\ \ = \ \ \ \ \ \ \ \ \ \\
=\!\!\! \int\limits_{\small\begin{array}{c} Ka=0\\ pK =
0\end{array}} DpDa\ e^{-(p\ +\ J_a K)(Qa\ +\ J_{p})\ +\ J_{a}KJ_p}\
\ = \int\limits_{\small\begin{array}{c} Ka=0\\ pK = 0\end{array}}
DpDa\ e^{-(p\ +\ J_a K)Q(a\ +\ KJ_{p}) +\ J_{a}KJ_p}\ \ \sim\ \
{\Huge e^{J_a K J_p}}
\end{split}
\end{equation}
Here\footnote{In these expressions we omit the pairings $\int$,
$STr$ and $<\ ,\ >$ in the action.} in the first equality we
inserted $\{K\ ,\ Q\}\ =\ 1$ in the space $V$ (\ see discussion
after (\ref{Lagr})\ ). In the second equality we explored the
definition of lagrangian sub-manifold $Ka\ =\ 0$ and extracted the
total square. In the next equality we again inserted $\{K\ ,\ Q\}\
=\ 1$ in front of the source $J_p$ and used that $(p\ +\ J_a K)K\ =\
0$ due to $K^2=0$ and the definition of lagrangian sub-manifold. The
last proportionality sign is obtained after the integration over $a$
and $p$. One can shift the integration variables by the terms $J_a
K$ and $K J_p$ because this shift is along the integration domain.
Thus the propagator in (\ref{Sdiagr}) is given by $K$.

 To find the amplitude in figure \ref{4-point} one should first
multiply two representatives
$\mathsf{A_1,A_2}\in\mathcal{H}(Q)\otimes Func(x)\otimes
T(\mathfrak{g})$ in the left vertex. Since both expressions are
closed, the result is the sum of a representative $\mathcal{H}(Q)$
and an exact expression\footnote{$g$ is the coupling constant.}
$$
g \mathsf{A_1\cdot A_2}\ =\ h\ +\ Q(\omega)
$$
where $h\in\mathcal{H}(Q)$ and $K\omega\ =\ 0$. At the next step one
should apply to this result operator $K$, which is responsible for
propagating quantum field. Strictly speaking operator $K$ is defined
only on the space $V$ and can not be applied to $\mathcal{H}(Q)$.
However we extend the domain  of $K$ and define this action as $K
\mathcal{H}(Q)\ =\ 0$. Thus, due to (\ref{K}), the result is
$$
K(g \mathsf{A_1\cdot A_2})\ =\ \omega
$$
To finish  evaluation of the diagram in figure \ref{4-point}, one
should multiply  the last in-line $\mathsf{A_3}$ by $\omega$ and
project the result onto $\mathcal{H}(Q)$. This is done by
calculation of the canonical pairing $<\mathsf{P},\
g\mathsf{A}_3\cdot\omega>\ =\ <\mathsf{P},\ g\mathsf{A}_3\cdot
K(g\mathsf{A}_1\cdot\mathsf{A}_2)>$ from the introduction. Such
projection is non-zero only if some part of expression $g^2
\mathsf{A_3}\cdot K( \mathsf{A_1\cdot A_2})$ is a representative of
$\mathcal{H}(Q)$. Otherwise this amplitude is equal to zero.

  Since  the theory (\ref{Fund}) contains Grassmann
variables it is important to fix convention of how one should
multiply the fields in each 3-valent vertex. We use the clockwise
rule for that. This means that one should multiply the fields  in
the clockwise direction, starting from the out-line. This convention
was illustrated in the example above: in the first vertex the order
of multiplication is $\mathsf{A_1}\cdot\mathsf{A_2}$ ( not
$\mathsf{A_2}\cdot\mathsf{A_1}$ ), in the second vertex
$\mathsf{A_3}\cdot\omega$ (not $\omega\cdot\mathsf{A_3}$).

\begin{figure}[h]
\centerline{\includegraphics[width=70mm]{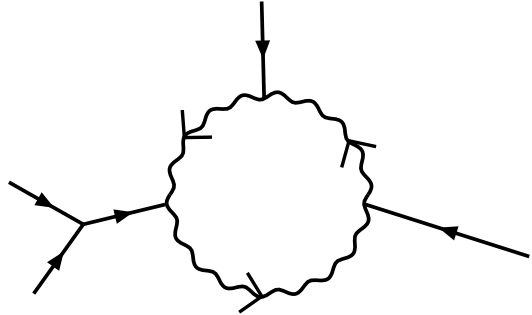}} \caption[
]{\label{loop}  {\footnotesize The only possible 1-loop diagram in
the theory (\ref{Sdiagr}).}}\noindent
\end{figure}

 By this remark we conclude our illustrative example with the 4-point
amplitude  and switch to evaluation of effective action for the
regular set (\ref{relations for CS }). In this example all
representatives of $\mathcal{H}(Q)$ have remarkable feature (see
explicit expressions (\ref{cohCS})\ ): $\#\lambda=\#\theta$ (number
of $\lambda$ $=$ number of $\theta$). Propagator $K\ =\ Q^{-1}\ =\
[\lambda_\alpha\frac{\partial}{\partial\theta_\alpha}]^{-1}$
 decreases $\#\lambda$ by $1$ and increases $\#\theta$ by $1$, violating
$\#\lambda\ =\ \#\theta$. Hence all the tree diagrams with
propagating quantum field (like the one in the figure \ref{4-point})
are forbidden (the canonical pairing is equal to zero). We have
explained above that diagram technique in the theory (\ref{Sdiagr})
is non-trivial only in one loop. It is obvious that for particular
example of operator $Q\ =\
\lambda_\alpha\frac{\partial}{\partial\theta_\alpha}$ in the kinetic
term even 1-loop diagrams are absent. One of several arguments for
this is that the only possible loop diagram is the one presented in
the figure (\ref{loop}).

\noindent This diagram can have only in-lines. Since each propagator
$Q^{-1}$ increases the number of $\theta$ by $1$, in-lines should
contribute some negative power of $\theta$, to have a chance to
close the loop. But there are no fields with negative degree of
$\theta$. Hence there are no loop diagrams at all.

 We conclude that in this particular example there are no loop
diagrams and there are no non-trivial tree diagrams, having
propagating quantum field. Hence, the only relevant diagrams are
those depicted in the figure~\ref{relevant CS}.

\newpage

\subsection{Calculation of Effective Action for the "Regular" Set (\ref{relations for CS })}
 A convenient way of evaluation for these diagrams is the
introduction of the  superfield
\begin{eqnarray}\label{superfield A}
\mathsf{A}\ \ \ \ =&c\ \ \ \ +\ \ \ \lambda_1\theta_1 A_1\ +\
\lambda_2\theta_2 A_2\ +\ \lambda_3\theta_3 A_3\ +\ & \\ & +\
\lambda_2\lambda_3\theta_2\theta_3 A_1^\ast\ -\
\lambda_1\lambda_3\theta_1\theta_3 A_2^\ast\ +\
\lambda_1\lambda_2\theta_1\theta_2 A_3^\ast\ +\ &
\!\!\!\!\lambda_1\lambda_2\lambda_3\theta_1\theta_2\theta_3\
c^{\ast}\nonumber
\end{eqnarray}
In this superfield all representatives of $\mathcal{H}(Q)$ found in
(\ref{cohCS}) are included. Letters $c, A_\mu, A^\ast_\mu, c^\ast\
\in Func(x)\otimes T(\mathfrak{g})$ stand for the fields which will
finally form effective action. Minus sign in front of
$\lambda_1\lambda_3\theta_1\theta_3$ is chosen for the future
convenience. From the definition of the superfield it is clear that
representatives of $\mathcal{H}(Q)$ play the role of polarizations
for the fields $c$, $A_\mu$, $A^\ast_\mu$, $c^\ast$. We would like
to emphasize that though in section 2 we used notation $A^\ast$ for
BV antifield to a field $A$, in the superfield (\ref{superfield A})
the field $A^\ast_\mu$ is not an antifield to $A_\mu$ and $c^\ast$
is not an antifield to $c$. They will become such fields only after
$Z_2$ reduction discussed in the section 2. While at the present
moment the antifields (see (\ref{superfield P}) below) are denoted
by tilde-sign, like $\widetilde{c}$, $\widetilde{A}_\mu$, etc.

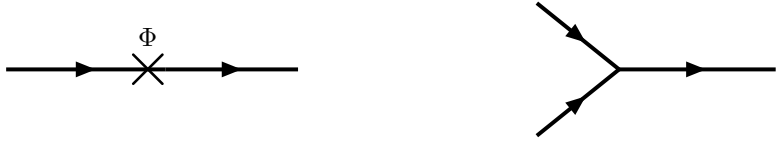
\begin{figure}[h]
\begin{center}
\begin{picture}(350,50)(0,275)

\ArrowLine(10,300)(70,300) \ArrowLine(70,300)(120,300)
\put(54,295){\Huge$\times$} \put(60,310){$\Phi$}
\ArrowLine(240,300)(300,300) \ArrowLine(210,275)(241,300)
\ArrowLine(210,325)(241,300)
\end{picture}  \\
\caption[ ]{\label{relevant CS}  {\footnotesize Relevant diagrams
for AKSZ-Chern-Simons theory, produced from the quadrics
(\ref{relations for CS }).}}
\end{center}
\end{figure}

\noindent To calculate the first diagram in the figure \ref{relevant
CS} we should apply  operator $\Phi$
$$
\Phi\ =\ \theta_\alpha\frac{\partial
f^{\mu}}{\partial\lambda_\alpha}\partial_\mu\ =\ \lambda_1\theta_1
\partial_1\ +\ \lambda_2\theta_2 \partial_2\ +\ \lambda_3\theta_3
\partial_3
$$
to the superfield $\mathsf{A}$, which stands for the in-line in this
diagram, and project the result onto cohomologies $\mathcal{H}(Q)$.
This is done by calculation of the canonical pairing $<\ ,\ >$ with
the anti-superfield\footnote{We emphasize that the pairs (BV field\
,\ BV antifield) in (\ref{superfield A}) and (\ref{superfield P})
are organized as follows: ($c$,\ $\widetilde{c}$), ($A_\mu$,
$\widetilde{A}_\mu$), ($A^\ast_\mu$,~$\widetilde{A}^\ast_\mu$),
($c^\ast$,~$\widetilde{c}^\ast$).}
\begin{eqnarray}\label{superfield P}
\mathsf{P}\ \ \ \ =&\widetilde{c}\ \ \ \ -\ \ \
\underline{\lambda_1\theta_1}\widetilde{A}_1\ -\
\underline{\lambda_2\theta_2}\widetilde{A}_2\ -\
\underline{\lambda_3\theta_3}\widetilde{A}_3\ +\ & \\ & +\
\underline{\lambda_2\lambda_3\theta_2\theta_3}\widetilde{A}^\ast_1\
-\
\underline{\lambda_1\lambda_3\theta_1\theta_3}\widetilde{A}^\ast_2\
+\
\underline{\lambda_1\lambda_2\theta_1\theta_2}\widetilde{A}^\ast_3\
-\ &
\!\!\!\!\underline{\lambda_1\lambda_2\lambda_3\theta_1\theta_2\theta_3}\
\widetilde{c}^\ast\nonumber
\end{eqnarray}
For instance, take the field $\lambda_1\theta_1 A_1$. One should
apply operator $\Phi$ to it:
$$
\Phi\big( \lambda_1\theta_1 A_1 \big)\ =\
\lambda_1\lambda_2\theta_2\theta_1\partial_2 A_1\ +\
\lambda_1\lambda_3\theta_3\theta_1\partial_3 A_1
$$
The result of projection onto $\mathcal{H}(Q)$ is given by:
\begin{equation}\nonumber
\begin{split}
<\mathsf{P},\ \lambda_1\lambda_2\theta_2\theta_1\partial_2 A_1\ +\
\lambda_1\lambda_3\theta_3\theta_1\partial_3 A_1>\ =\
<\underline{\lambda_1\lambda_2\theta_1\theta_2}
\widetilde{A}^\ast_3,\
\lambda_1\lambda_2\theta_2\theta_1\partial_2A_1>\ +\\ +\
<-\underline{\lambda_1\lambda_3\theta_1\theta_3}\widetilde{A}^\ast_2,\
\lambda_1\lambda_3\theta_3\theta_1\partial_3A_1>\ =\
-\widetilde{A}^\ast_3\partial_2A_1\ +\
\widetilde{A}^\ast_2\partial_3A_1
\end{split}
\end{equation}
Conducting such manipulations for each component field of the
superfield $\mathsf{A}$ one can obtain the first line of the
effective action (\ref{EFF-PCS}). Evaluation of the second diagram
of figure \ref{relevant CS} goes in a similar way: one should
multiply two superfields $\mathsf{A}$ corresponding to the in-lines
and project the result onto the out-line  - the superfields
$\mathsf{P}$. The sum of contributions of these two diagrams is
given by\footnote{\label{internal parity}It is important to remember
the internal parities of the fields (see the table below). For
example, the ghost field $c$ has negative parity, hence anticommutes
with $\theta_\alpha$:\ \ \ $c\lambda_1\theta_1A_1\ +\
\lambda_1\theta_1A_1c\ =\ \lambda_1\theta_1(A_1c\ -\ cA_1)$. This
expression contributes into the third term of the second line in
(\ref{EFF-PCS}).}
\begin{equation}\label{EFF-PCS}
\begin{split} S^{eff}\ \ \ =\ \ \ \int d^3x\ STr
\Big(\
 -\ \widetilde{A}_{\mu}\partial_\mu c\ \ -\ \
\widetilde{c}^\ast
\partial_\mu A_\mu^\ast\ \ +\ \
\varepsilon^{\mu\nu\lambda}\widetilde{A}^\ast_\mu
\partial_\nu A_\lambda\ \ +\ \ \ \ \ \ \ \ \ \ \    \\
+\ g\varepsilon^{\mu\nu\lambda} \widetilde{A}^\ast_\mu A_\nu
A_\lambda\ +\ g\widetilde{c} cc\ +\ g\widetilde{A}_\mu[c,\ A_\mu]\
+\ g\widetilde{A}^\ast_\mu\{c,\ A_\mu^\ast\}\ +\
g\widetilde{c}^\ast\ [c,\ c^\ast]\ +\ g\widetilde{c}^\ast\
[A_\mu^\ast,\ A_\mu] \Big)
\end{split}
\end{equation}
Since we have started from BV action (\ref{Sdiagr}) and integrated
over the lagrangian sub-manifold (\ref{Lagr}), effective action
should be again BV version of some theory. It happens that
(\ref{EFF-PCS}) is nothing but AKSZ-type BV action for Chern-Simons
theory (\ref{PCS}) (see for example \cite{Tonin}). To see this, one
should expand the fields $\Psi\ =\ \Psi^{(0)}\ +\ \Psi^{(1)}\ +\
\Psi^{(2)}\ +\ \Psi^{(3)}$ and $\Psi^\ast\ =\ \Psi^{\ast(0)}\ +\
\Psi^{\ast(1)}\ +\ \Psi^{\ast(2)}\ + \ \Psi^{\ast(3)}$ in the action
(\ref{PCS}). The result is
\begin{equation}\label{PCS expanded}
\begin{split}
S^{AKSZ-CS}\ =\ \int STr \Big( \Psi^{\ast(0)}d\Psi^{(2)}\ +\
\Psi^{\ast(2)}d\Psi^{(0)}\ +\ \Psi^{\ast(1)}d\Psi^{(1)}\ +\
g\Psi^{\ast(3)}\Psi^{(0)}\Psi^{(0)}\ +\
g\Psi^{\ast(2)}\{\Psi^{(0)},\Psi^{(1)}\}\ +\ \\ +\
g\Psi^{\ast(1)}\{\Psi^{(0)},\Psi^{(2)}\}\ +\
g\Psi^{\ast(0)}\{\Psi^{(0)},\Psi^{(3)}\}\ +\
g\Psi^{\ast(1)}\Psi^{(1)}\Psi^{(1)}\ +\
g\Psi^{\ast(0)}\{\Psi^{(1)},\Psi^{(2)}\} \Big)
\end{split}
\end{equation}
Action (\ref{PCS expanded}) reproduces effective action
(\ref{EFF-PCS}) under the identification
$${
\begin{array}{cccc}
\!\!\!\!\Psi^{(0)}\ =\ c\ \ \  &\ \ \  \Psi^{(1)}\ =\ dx^\mu A_\mu\
\ \ &\ \ \ \Psi^{(2)}\ =\ \frac{1}{2}\ dx^\mu\wedge dx^\nu\
\varepsilon_{\mu\nu\lambda}\ A_\lambda^{\ast} \ \ \ &\ \ \
\Psi^{(3)}\ =\ \frac{1}{6}\ dx^\mu\wedge dx^\nu\wedge dx^\lambda\ \varepsilon_{\mu\nu\lambda}\  c^{\ast}\\  \\
\Psi^{\ast(0)}\ =\ \widetilde{c}^\ast\ \ \  &\ \ \ \Psi^{\ast(1)}\
=\ dx^\mu \widetilde{A}^\ast_\mu\ \ \  &\ \ \ \Psi^{\ast(2)}\ =\
\frac{1}{2}\ dx^\mu\wedge dx^\nu\ \varepsilon_{\mu\nu\lambda}\
\widetilde{A}_\lambda \ \ \ &\ \ \ \Psi^{\ast(3)}\ =\ \frac{1}{6}\
dx^\mu\wedge dx^\nu\wedge dx^\lambda\ \varepsilon_{\mu\nu\lambda}\
\widetilde{c}
\end{array}
}$$ In this calculation one should take into account the parities of
the fields:
$$
\begin{tabular}{|c|c|}
\hline
 {\bf even} & {\bf odd} \\
\hline & \\
$ A_\mu\ \ c^\ast\ \ \ \ \widetilde{A}_\mu^\ast\ \ \widetilde{c} $& $ A_\mu^\ast\ \ c\ \ \ \ \widetilde{A}_\mu\ \ \widetilde{c}^\ast $ \\
&\\ \hline
\end{tabular}
$$
and that odd fields anticommute with the generators of exterior
algebra $dx^\mu$. Hence, in most of terms  in (\ref{PCS expanded})
anticommutators transform into commutators.

  Thus, we have demonstrated that AKSZ version of Chern-Simons theory
can be obtained as effective action from the theory (\ref{Fund}). It
is instructive to note that the action (\ref{EFF-PCS}) transforms
into normal BV action for Chern-Simons gauge theory under the
reduction over half of degrees of freedom
$$
\begin{array}{cc}
\widetilde{A}^\ast_\mu\ =\ A_\mu\ \ \ \ \  & \widetilde{c}^\ast\ =\ c\\
\widetilde{A}_\mu\ =\ A_\mu^\ast\ \ \ \ \  & \widetilde{c}\ =\
c^\ast
\end{array}
$$
This $Z_2$ reduction was mentioned in the section 2, see also figure
\ref{fig1}. It is possible to do such reduction in case the
dimension of space-time is odd. After this reduction the action
(\ref{EFF-PCS}) transforms to
\begin{equation}\nonumber
\begin{split} S^{CS-gauge}\ \ \ =\ \ \ \int d^3x\ STr
\Big(\ -\ 2A^\ast_{\mu}\partial_\mu c\ +\ 3g A^\ast_\mu[c,\ A_\mu]\
+\ 3 g\ \! c^\ast c c\ +\
\varepsilon^{\mu\nu\lambda}A_\mu\partial_\nu A_\lambda\ +\
g\varepsilon^{\mu\nu\lambda}A_\mu A_\nu A_\lambda\Big)
\end{split}
\end{equation}
which is BV action of Chern-Simons gauge theory under redefinitions
$g\longrightarrow\frac{2}{3}g$,
$c^\ast\longrightarrow\frac{1}{2}c^\ast$ and
$A_\mu^\ast\longrightarrow\frac{1}{2}A_\mu^\ast$. In this action the
field $A_\mu$ is the gauge field, $A^\ast_\mu$ is BV anti-gauge
field, $c$ is the ghost field and $c^\ast$ is BV anti-ghost field.

\subsection{Generalization for arbitrary "regular" system of constraints }
  Thus, we have demonstrated that the theory (\ref{EFF-PCS}) arises
as effective action for particular choice of constraints
(\ref{relations for CS }). It happens that this result is in fact
more general. The same effective action can be obtained for
arbitrary "{\bf regular}" set of constraints. To see this, note that
in case of {\bf "regular"} quadrics the first-level relations can be
written as
$$
G^{(1)\ \mu_{1}}\ =\ \varepsilon^{\mu_{1}\mu_{2}}f_{\mu_2}
$$
where $\varepsilon^{\mu_1\mu_2}$ is an arbitrary antisymmetric
matrix. It is convenient to choose a basis in the space of such
matrices. The basis elements are denoted by two indices $\alpha_1$
and $\alpha_2\ \ \ $: $\varepsilon^{\mu_1\mu_2}_{\alpha_1\
\alpha_2}$ - the matrix which has $1$ at the intersection of
$\alpha_1$-th line and $\alpha_2$-th column and $-1$ at the
$\alpha_1$-th column and $\alpha_2$-th line. Explicit expression for
such matrix is
$$
\varepsilon^{\mu_1\mu_2}_{\alpha_1\ \alpha_2}\ =\
\big(\delta^{\mu_1}_{\alpha_1}\delta^{\mu_2}_{\alpha_2}\ -\
\delta^{\mu_1}_{\alpha_2}\delta^{\mu_2}_{\alpha_1} \big)
$$
Thus, in {\bf "regular"} case the first-level relations are:
$$
G^{(1)\ \mu_1}_{\alpha_1\ \alpha_2}\ =\
\big(\delta^{\mu_1}_{\alpha_1}\delta^{\mu_2}_{\alpha_2}\ -\
\delta^{\mu_1}_{\alpha_2}\delta^{\mu_2}_{\alpha_1} \big) f_{\mu_2}
$$
They obviously satisfy $G^{(1)\ \mu_1}_{\alpha_1\
\alpha_2}f_{\mu_1}\ =\ 0$. Explicit expressions for the second-level
relations are
$$
G^{(2)\ \alpha_1\ \alpha_2}_{\beta_1\ \beta_2\ \beta_3}\ =\
\delta^{\alpha_1}_{[\beta_1}\delta^{\alpha_2}_{\beta_2}\delta^{\alpha_3}_{\beta_3]}\
f_{\alpha_3}
$$
where one should make antisymmetrization over the indices
$\beta_1$,$\beta_2$,$\beta_3$. Expressions for the n-th level
relations are
$$
G^{(n)\ \ \alpha_1,.....,\alpha_n}_{\ \ \ \
\beta_1,.....,\beta_{n+1}}\ =\
\delta^{\alpha_1}_{[\beta_1}.....\delta^{\alpha_{n+1}}_{\beta_{n+1}]}\
f_{\alpha_{n+1}}
$$
To extract cohomologies from this system of relations one should
apply operator $\widetilde{D}$ to them according to (\ref{coh}).
Straightforward calculation for the first-level relations gives
$$
\widetilde{D}G^{(1)\ \!\beta_1}_{\alpha_1\
\alpha_2}\widetilde{D}f_{\beta_1}\ =\
2\widetilde{D}f_{\alpha_2}\widetilde{D}f_{\alpha_1}
$$
Similar computation for higher-order relations demonstrate that the
algebra of cohomologies (\ref{coh}) is isomorphic to exterior
algebra
$\Lambda^{\bullet}\{\widetilde{D}f_1,.....,\widetilde{D}f_N\}$ of
anticommuting  generators $\widetilde{D}f_1,.....,\widetilde{D}f_N$.
Hence, we conclude that for {\bf "regular"} system of quadrics the
algebra of cohomologies is universal and does not depend on explicit
expressions of quadrics $f^\mu(\lambda)$. The same is true for the
effective action. Since $\mathcal{H}(Q)\ =\
\Lambda^{\bullet}\{\widetilde{D}f_1,.....,\widetilde{D}f_N\}$, each
representative has  $\#\lambda=\#\theta$ (degree of $\lambda_\alpha$
is the same as degree of $\theta_\alpha$). Using the same argument
as in the end of section 5.2 we conclude that the only relevant
diagrams for arbitrary {\bf "regular"} system are those of figure
\ref{relevant CS}. Since algebra of cohomologies $\mathcal{H}(Q)$ is
universal and diagrams are universal, effective action is also
universal. This effective action is AKSZ-Chern-Simons theory in $N$
dimensions. (This  was demonstrated in section 5.3 for $N=3$.
Generalization to higher $N$ is straightforward.) This theory is
well defined for arbitrary (including even) value of $N$.

  Important observation is that almost all possible constraints
are {\bf "regular"} (this is true if the number of quadrics is less
then the number of $\lambda_\alpha$: $N\leq K$). Indeed, suppose we
have written without thinking some random set of quadrics:
$${
\begin{array}{l}
f_1\ = \ 3.14159\lambda_1^2\ +\ 2\lambda_2\lambda_3\ +\ \ln5
\lambda_1\lambda_2\\
f_2\ =\ 10\lambda_1^2\ +\ \lambda_2^2\ +\
\sqrt{2}\lambda_1\lambda_2\\
f_{3}\ =\ \lambda_1\lambda_3
\end{array}}
$$
Most probably, if this set is enough complicated,  there are only
trivial relations, obtained by antisymmetrization of
$f^{\mu}(\lambda)$. Hence, the set is "{\bf regular}". Thus, if we
take a random point in the space of quadrics, we, most probably,
obtain a {\bf "regular"} system, hence the theory of
AKSZ-Chern-Simons as effective action.

  Concluding this section we see that effective field theory is in
one to one correspondence with the coefficients of quadrics
$f^{\mu}(\lambda)$. In the space of these coefficients "{\bf
regular}" points correspond to AKSZ-Chern-Simons theories. Nearly
all points in this space are "{\bf regular}". However, there  is a
small, but still infinite, number of "{\bf singular}" points. In
this points some non-trivial diagrams survive, hence non-trivial
theories appear. In the next section we study one example of such
{\bf "singular"} point. We demonstrate that effective action
(\ref{BVVintegral}) in this point  reproduces  BV version of two
dimensional model, which contains a gauge field, scalars and some
number of fermions.

\vspace{0.7cm}

\section{The Gauge Model}\label{Gauge model chapter}
 This section presents our main result in the present paper. We
demonstrate  that in the space $O[\theta_\alpha\ \! ,\ \!
\lambda_\alpha\ |\ f^{\mu}(\lambda)]$ generated by 4 variables
$\lambda_\alpha$ and $\theta_\alpha$ $(\alpha=1..4)$ there exists a
{\bf "singular"} set of 5 constraints $f^\mu(\lambda)$:
\begin{eqnarray}\label{singular constraints}
f_1\ =\ \lambda_1\lambda_2\nonumber\\
f_2\ =\ \lambda_2\lambda_3\nonumber\\
f_3\ =\ \lambda_3\lambda_4\\
f_4\ \ \ =\ \ \lambda_1^2\nonumber\\
f_5\ \ \ =\ \ \lambda_4^2\nonumber
\end{eqnarray}
such that effective BV action for the theory  (\ref{Fund})
represents the two-dimensional  gauge model.

  The system of quadrics (\ref{singular constraints}) is invariant
under the discrete $Z_2$ symmetry:
\begin{equation}\label{symmetry}
{\begin{array}{c}
\lambda_1\ \leftrightarrow\ \lambda_4\\
\lambda_2\ \leftrightarrow\ \lambda_3
\end{array}}\ \ \ \ \ \ \ \ \
{\begin{array}{c}
f_1\ \leftrightarrow \ f_3\\
f_2\ \leftrightarrow\  f_2\\
f_4\ \leftrightarrow\  f_5
\end{array}}
\end{equation}
which after  calculation of effective action becomes the symmetry
between the left and right fields.

  We begin analysis of the system (\ref{singular constraints}) from
evaluation of all representatives of $\mathcal{H}(Q)$ along the
lines of section~\ref{Cohomology evaluation}. The first-level
relations, written in the basis $(f_1,\ f_2,\ f_3,\ f_4,\ f_5)$ are:
$$
\begin{array}{ccccccc}
G^{(1)}_1 & = & \lambda_3 & -\lambda_1 & 0 & 0 & 0 \\
G^{(1)}_2 & = & \lambda_1 & 0 & 0 & -\lambda_2 & 0 \\
G^{(1)}_3 & = & 0 & \lambda_4 & -\lambda_2 & 0 & 0 \\
G^{(1)}_4 & = & 0 & 0 & \lambda_4 & 0 & -\lambda_3 \\
G^{(1)}_5 & = & \lambda_4^2 & 0 & 0 & 0 & -\lambda_1\lambda_2 \\
G^{(1)}_6 & = & 0 & 0 & \lambda_1^2 & -\lambda_3\lambda_4 & 0 \\
G^{(1)}_7 & = & 0 & 0 & 0 & \lambda_4^2 & -\lambda_1^2 \\
\end{array}
$$
Note, that relations $G^{(1)}_5,\ G^{(1)}_6,\ G^{(1)}_7$ are trivial
according to the definition above the example in section 4.1.
However there are additional relations $G^{(1)}_1,\ G^{(1)}_2,\
G^{(1)}_3,\ G^{(1)}_4$ which are non-trivial. Hence, the system of
quadrics (\ref{singular constraints}) is {\bf "singular"}.

 Secondary relations in the basis $(G^{(1)}_1,\ G^{(1)}_2,\
G^{(1)}_3,\ G^{(1)}_4,\ G^{(1)}_5,\ G^{(1)}_6,\ G^{(1)}_7)$ are
$$
\begin{array}{ccccccccc}
G^{(2)}_1  & = & 0 & \lambda_4^2 & 0 & 0 & -\lambda_1 & 0 & \lambda_2 \\
G^{(2)}_2  & = & 0 & 0 & 0 & \lambda_1^2 & 0 & -\lambda_4 & -\lambda_3 \\
G^{(2)}_3  & = & \lambda_4^2 & 0 & \lambda_1\lambda_4 & \lambda_1\lambda_2 & -\lambda_3 & 0 & 0 \\
G^{(2)}_4  & = & \lambda_1\lambda_4 & -\lambda_3\lambda_4 & \lambda_1^2 & 0 & 0 & \lambda_2 & 0 \\
\end{array}
$$
There is only one third-level relation $G^{(3)}$
$$
\lambda_3G^{(2)}_1\ +\ \lambda_2G^{(2)}_2\ -\ \lambda_1G^{(2)}_3\ +\
\lambda_4G^{(2)}_4\ =\ 0
$$
which completes the tower of relations. It is straightforward to
obtain all representatives of $\mathcal{H}(Q)$ by applying operator
$\widetilde{D}$ to these relations. The result is presented in the
first column of the table on the next page\footnote{Though the
calculation is straightforward, we give an example here, which also
clarifies one subtlety in this table. Consider the cohomology,
generated by the third second-level relation $G^{(2)}_3$:
\begin{equation}\nonumber
\begin{split}
\widetilde{D}\Bigg[\widetilde{D}\Big(\widetilde{D}\big(G^{(2)\
\!A}_3\big)\ G^{(1)\ \! \mu}_A\ \Big)\ f_\mu\Bigg]\ =\
\frac{16}{30}\lambda_2\lambda_4\theta_1\theta_3\theta_4\ +\
\frac{13}{30}\lambda_1\lambda_4\theta_2\theta_3\theta_4
\end{split}
\end{equation}
This expression is different from the polarization of the field
$\varphi_7$ in the table. However, since cohomologies are
equivalence classes, one has a freedom to add the exact term
$-\frac{3}{60}Q(\lambda_4\theta_2\theta_1\theta_3\theta_4)$ to this
result to obtain the expression proportional to that in the table.
Addition of such exact expression is done only for convenience. It
is allowed to add such exact terms to the polarizations of fields
which do not contribute to the diagrams with propagator. For
diagrams with propagator, addition of exact terms can change the
result for the effective action. Hence one should use precisely the
representatives given by expressions (\ref{coh}).}.

  In the previous section we used the notation of the superfield as
a tool for diagram calculation. Since the number of cohomologies for
the set (\ref{singular constraints}) is large, it is more convenient
to present this superfield in the form of the table. Second and
third columns in this table stand for notation of component field
and antifield respectively. In the first column the
$\lambda$-$\theta$ structure (polarization) for these component
fields is written. By horizontal lines all representatives are
divided in groups having equal number of $\lambda$ and $\theta$.

It should be emphasized here that in the previous chapters we
denoted the BV antifield to a field $A$ as $A^\ast$. In the present
section we denote BV antifield to a field $A$ as $\widetilde{A}$.
This is done only for convenience, because as will be clear in the
end of the section some of the fields $\widetilde{A}$ will
contribute into the classical part of the effective BV action.

$$
\begin{array}{|c|c|c|}
\hline
$Polarization$  & $Field$ & $Antifield$ \\
\hline
 & &\\
1  & c & \widetilde{c} \\
 & &\\
 \hline
 & &\\
\lambda_1\theta_2\ +\ \lambda_2\theta_1  & \gamma_+ & \widetilde{\gamma}_+ \\
\lambda_2\theta_3\ +\ \lambda_3\theta_2  & \varphi & \widetilde{\varphi} \\
\lambda_3\theta_4\ +\ \lambda_4\theta_3  & \gamma_- & \widetilde{\gamma}_- \\
\lambda_1\theta_1  & A_+ & \widetilde{A}_+ \\
\lambda_4\theta_4   & A_- &  \widetilde{A}_- \\
 & &\\
\hline
 & &\\
\lambda_1\theta_1\theta_2  & \psi_+ & \widetilde{\psi}_+ \\
\lambda_4\theta_4\theta_3  & \psi_- &  \widetilde{\psi}_-\\
\frac{1}{3}(\lambda_1\theta_3\theta_2\ +\ \lambda_3\theta_2\theta_1\ +\ 2\lambda_2\theta_3\theta_1)  & \chi_+ & \widetilde{\chi}_+ \\
\frac{1}{3}(\lambda_2\theta_3\theta_4\ +\ \lambda_4\theta_2\theta_3\ +\ 2\lambda_3\theta_2\theta_4)  & \chi_- & \widetilde{\chi}_- \\
 & &\\
\hline
 & &\\
\lambda_1\lambda_4\theta_1\theta_4  & \varphi_1 & \widetilde{\varphi}_1 \\
\lambda_1\lambda_4\theta_4\theta_2\ +\ \lambda_2\lambda_4\theta_4\theta_1  & \varphi_2 & \widetilde{\varphi}_2 \\
\lambda_1\lambda_3\theta_1\theta_4\ +\ \lambda_1\lambda_4\theta_1\theta_3  & \varphi_3 & \widetilde{\varphi}_3 \\
 & &\\
\hline
 & &\\
\lambda_1\lambda_4\theta_1\theta_4\theta_3  & \varphi_4 & \widetilde{\varphi}_4 \\
\lambda_1\lambda_4\theta_1\theta_2\theta_4  & \varphi_5 & \widetilde{\varphi}_5 \\
\lambda_1\lambda_3\theta_1\theta_2\theta_4\ +\ \lambda_1\lambda_4\theta_1\theta_2\theta_3  & \varphi_6 & \widetilde{\varphi}_6 \\
\lambda_2\lambda_4\theta_1\theta_4\theta_3\ +\ \lambda_1\lambda_4\theta_2\theta_4\theta_3  & \varphi_7 & \widetilde{\varphi}_7 \\
 & &\\
\hline
 & &\\
\lambda_1\lambda_4\theta_1\theta_2\theta_3\theta_4  & \varphi_8 & \widetilde{\varphi}_8 \\
 & &\\
\hline
\end{array}
$$
It is obvious that (\ref{symmetry}) is a symmetry of fields in the
table. For example, under this symmetry $A_+\ \leftrightarrow \ A_-\
$, $\ \chi_+\ \leftrightarrow \ \chi_-\ $, $\ \varphi\
\leftrightarrow \ \varphi\ $ and so on for all 18 fields of the
table. Most of fields in this table are doublets under this
symmetry, however there are singlets, like $c$, $\varphi$,
$\varphi_1$, $\varphi_8$.

  Operator $\Phi$, which is the second term in (\ref{Qtot}), for this model is given by
$$\begin{array}{c} \Phi\ =\ (\lambda_1\theta_2\ +\
\lambda_2\theta_1)\partial_1\ +\ (\lambda_2\theta_3\ +\
\lambda_3\theta_2)\partial_2\ +\ (\lambda_3\theta_4\ +\
\lambda_4\theta_3)\partial_3\ +\ 2\lambda_1\theta_1\partial_4\ +\
2\lambda_4\theta_4\partial_5\ =\ \\ \\=\
2\lambda_1\theta_1\partial_+\ +\ 2\lambda_4\theta_4\partial_-
\end{array}$$
Note, that in the second equality we use the freedom of choosing the
space $Func(x)$ in the definition of (\ref{Fund}). We choose this
space in such a way that all the fields in our model depend only on
$x^4$ and $x^5$ and do not depend on $x^1$,$x^2$,$x^3$, making a
reduction from 5 dimensional theory to two dimensions. These
derivatives are denoted by $\partial_4=\partial_+$ and
$\partial_5=\partial_-$.

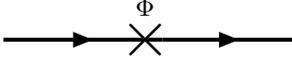
\begin{figure}[h]
\begin{center}
\begin{picture}(250,70)(-50,270)
\ArrowLine(10,300)(70,300) \ArrowLine(70,300)(120,300)
\put(54,295){\Huge$\times$} \put(60,310){$\Phi$}
\end{picture}  \\
\caption[ ]{\label{figS1}  {\footnotesize Kinetic term.}}
\end{center}
\end{figure}

\begin{figure}[h]
\begin{center}
\begin{picture}(350,70)(90,270)
\ArrowLine(240,300)(300,300) \ArrowLine(210,275)(241,300)
\ArrowLine(210,325)(241,300)
\end{picture}  \\
\caption[ ]{\label{figS2}  {\footnotesize 3-valent vertex.}}
\end{center}
\end{figure}
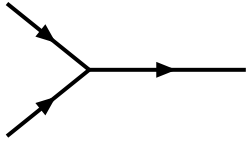

\begin{figure}[h]
\begin{center}
\begin{picture}(250,35)(-36,288)
\ArrowLine(10,300)(70,300) \Photon(70,300)(120,300){1}{4}
\put(60,294.9){\Huge$\times$} \put(66,310){$\Phi$}
\put(110,294.9){\Huge$\times$} \put(116,310){$\Phi$}
\ArrowLine(120,300)(170,300)\Line(88,306)(95,299)
\Line(88,294)(95,299)
\end{picture}  \\
\caption[ ]{\label{figS3}  {\footnotesize Kinetic term in the gauge
sector.}}
\end{center}
\end{figure}
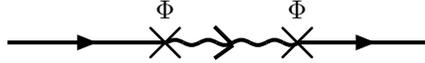

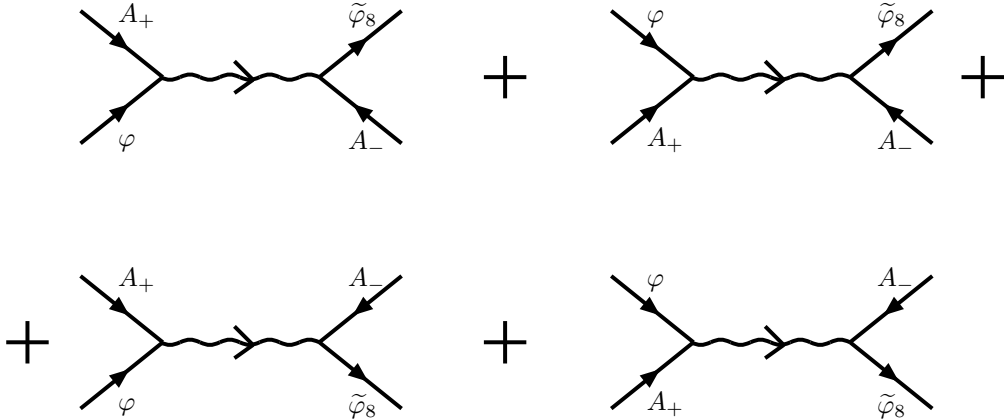
\begin{figure}[h]
\begin{center}
\begin{picture}(600,150)(125,180)
\Photon(220,300)(280,300){-1}{4} \ArrowLine(190,275)(221,300)
\ArrowLine(190,325)(221,300) \ArrowLine(280,300)(311, 325)
\ArrowLine(311,275)(280,300)\put(205,275){$\varphi$}
\put(205,322){$A_{+}$}\put(292,274){$A_{-}$}
\put(292,322){$\widetilde{\varphi}_8$}\Line(248,306)(255,299)
\Line(248,294)(255,299)

\Photon(420,300)(480,300){-1}{4} \ArrowLine(390,275)(421,300)
\ArrowLine(390,325)(421,300) \ArrowLine(480,300)(511, 325)
\ArrowLine(511,275)(480,300)\put(405,275){$A_+$}
\put(405,322){$\varphi$}\put(492,274){$A_{-}$}
\put(492,322){$\widetilde{\varphi}_8$}\Line(448,306)(455,299)
\Line(448,294)(455,299)

\Photon(220,200)(280,200){-1}{4} \ArrowLine(190,175)(221,200)
\ArrowLine(190,225)(221,200) \ArrowLine(311, 225)(280,200)
\ArrowLine(280,200)(311,175)\put(205,175){$\varphi$}
\put(205,222){$A_{+}$}\put(292,174){$\widetilde{\varphi}_8$}
\put(292,222){$A_{-}$}\Line(248,206)(255,199)
\Line(248,194)(255,199)

\Photon(420,200)(480,200){-1}{4} \ArrowLine(390,175)(421,200)
\ArrowLine(390,225)(421,200) \ArrowLine(511, 225)(480,200)
\ArrowLine(480,200)(511,175)\put(405,175){$A_+$}
\put(405,222){$\varphi$}\put(492,174){$\widetilde{\varphi}_8$}
\put(492,222){$A_{-}$}\Line(448,206)(455,199)
\Line(448,194)(455,199)

\Line(350,192)(350,208)\Line(342,200)(358,200)
\Line(350,292)(350,308)\Line(342,300)(358,300)
\Line(170,192)(170,208)\Line(162,200)(178,200)
\Line(530,292)(530,308)\Line(522,300)(538,300)
\end{picture}  \\
\caption[ ]{\label{figS4}  {\footnotesize Quartic contribution into
effective action for the gauge sector.}}
\end{center}
\end{figure}

\begin{figure}[h]
\begin{center}
\begin{picture}(600,70)(140,270)
\Photon(200,300)(250,300){-1}{4} \ArrowLine(170,275)(201,300)
\ArrowLine(170,325)(201,300)\ArrowLine(250,300)(285,300)
\put(240,295){\Huge$\times$}\put(246,310){$\Phi$}
\Line(538,306)(545,300) \Line(538,294)(545,300)

\Photon(370,300)(410,300){-1}{4} \ArrowLine(410,300)(441, 325)
\ArrowLine(441,275)(410,300)\ArrowLine(333,300)(370,300)
\put(384,274){$A_{+}$ and $A_-$}
\put(422,322){$\widetilde{\varphi}_8$}
\put(360,295){\Huge$\times$}\put(366,310){$\Phi$}
\Line(310,292)(310,308)\Line(302,300)(318,300)
\Line(460,292)(460,308)\Line(452,300)(468,300)
\put(345,288){$\varphi$}\Line(388,306)(395,300)
\Line(388,294)(395,300)

\Photon(520,300)(560,300){-1}{4} \ArrowLine(591, 325)(560,300)
\ArrowLine(560,300)(591,275)\ArrowLine(483,300)(520,300)
\put(572,274){$\widetilde{\varphi}_8$}
\put(510,295){\Huge$\times$}\put(516,310){$\Phi$}
\put(533,322){$A_{+}$ and $A_-$}
\put(500,288){$\varphi$}\Line(218,306)(225,300)
\Line(218,294)(225,300)
\end{picture}  \\
\caption[ ]{\label{figS5}  {\footnotesize Cubic contribution into
effective action for the gauge sector.}}
\end{center}
\end{figure}
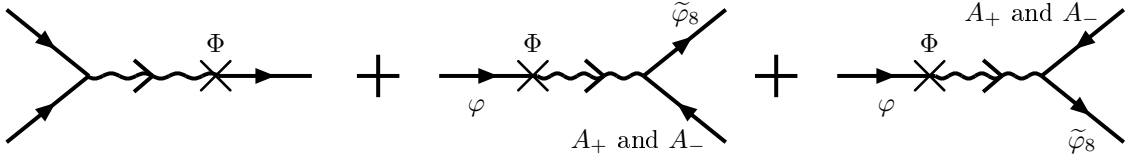

  For the theory (\ref{singular constraints}) all non-trivial
diagrams are  shown in the figures \ref{figS1} - \ref{figS5}. We
start from the simplest one in the figure \ref{figS1}.
To calculate this diagram one should take a representative from the
table, apply operator $\Phi$ to it and project the result onto
cohomologies $\mathcal{H}(Q)$  according to the canonical pairing
$<\ ,\
>$  from the introduction. For instance take the
representative $\lambda_1\theta_1A_+$
$$
\Phi(\lambda_1\theta_1A_+)\ =\
2\lambda_1\lambda_4\theta_4\theta_1\partial_-A_+\
\Big|_{Projecting}\!\!\!\longrightarrow\ \ \
-2\widetilde{\varphi}_1\partial_-A_+
$$
In the last equality we projected the result onto component field
$\lambda_1\lambda_4\theta_1\theta_4\widetilde{\varphi}_1$ of the
superfield $\mathsf{P}$ which includes all the fields from the last
column of the table:\ \
$<\underline{\lambda_1\lambda_4\theta_1\theta_4}\widetilde{\varphi}_1,\
2\lambda_1\lambda_4\theta_4\theta_1\partial_-A_+>\ =\
-2\widetilde{\varphi}_1\partial_-A_+$. This is one term in the
effective action. The whole contribution of the diagram in figure
\ref{figS1} is
\begin{equation}\nonumber
\begin{split}
S^{(1)}\ =\ \int\ STr\Big( 2\widetilde{A}_+\partial_+ c\ +\
2\widetilde{A}_- \partial_- c\ +\
2\widetilde{\varphi}_3\partial_+\gamma_-\ +\
2\widetilde{\varphi}_2\partial_-\gamma_+\ +\
2\widetilde{\varphi}_1(\partial_+A_-\ -\ \partial_-A_+)\ +\ \\
\\ +\ 2\widetilde{\varphi}_4\partial_+\psi_-\ +\
2\widetilde{\varphi}_5\partial_-\psi_+\ +\
\widetilde{\varphi}_6\partial_+\chi_-\ +\
\widetilde{\varphi}_7\partial_-\chi_+\Big)
\end{split}
\end{equation}
were integral now denotes usual integration over two space-time
dimensions.

\noindent Next diagram is the one in the figure \ref{figS2}.
Again, as an illustrative example consider the fields $A_-$ and
$A_+$. Using the clockwise  rule of section 5.2 the result is given
by
$$
(\lambda_1\theta_1A_+)(\lambda_4\theta_4A_-)\ =\
\lambda_1\lambda_4\theta_1\theta_4A_+A_-\Big|_{Projecting}\!\!\!\longrightarrow\
\  \widetilde{\varphi}_1A_+A_-
$$
There is another contribution to this diagram from the exchange of
the in-legs:
$$
(\lambda_4\theta_4A_-)(\lambda_1\theta_1A_+)\ =\
-\lambda_1\lambda_4\theta_1\theta_4A_-A_+\Big|_{Projecting}\!\!\!\longrightarrow\
\  -\widetilde{\varphi}_1A_-A_+
$$
Hence, the contribution into effective Lagrangian is $L\ =\ Tr\
g\widetilde{\varphi}_1[A_+,A_-]$ the commutator of fields $A_+$ and
$A_-$. The total result for all the fields is:
$${
\begin{split}
S^{(2)}\ =\ \int\ STr\  g\Bigg( \widetilde{\varphi}_1[A_+,A_-]\ +\
\widetilde{\varphi}_2[A_-,\gamma_+]\ +\
\widetilde{\varphi}_3[A_+,\gamma_-]\ +\
\widetilde{\varphi}_4[A_+,\psi_-]\ +\
\widetilde{\varphi}_5[A_-,\psi_+]\ +\  \\ \\ +\
\widetilde{\varphi}_6[\gamma_-,\psi_+]\ +\
\frac{1}{2}\widetilde{\varphi}_6[A_+,\chi_-]\ +\
\widetilde{\varphi}_7[\gamma_+,\psi_-]\ +\
\frac{1}{2}\widetilde{\varphi}_7[A_-,\chi_+]\ -\
\widetilde{\varphi}_8\{\psi_+,\psi_-\}\ +\\ \\ +\ \widetilde{c}cc\
+\ \widetilde{\gamma}_+[\gamma_+,c]\ +\
\widetilde{\varphi}[\varphi,c]\ +\ \widetilde{\gamma}_-[\gamma_-,c]\
+\ \widetilde{A}_+[A_+,c]\ +\ ......\ +\
\widetilde{\varphi}_8\{\varphi_8,c\}\Bigg)
\end{split}}
$$
In the  last line the standard terms, resulting from the
anti-commutator with the ghost field $c$ are written. Such terms are
non-vanishing for all 18 fields of the table. They are denoted by
multi-dot sign. Again, we emphasize that it is important to take
into account internal parities of the fields (see footnote
\ref{internal parity}). At this step one can have an impression that
the notion of internal parities is a sort of guess that one can make
looking at the final answer for the effective action. In fact this
is not the case. Internal parities can be directly extracted from
the homological analysis. The point is that all the fields in the
table above are divided by horizontal lines into blocs having
certain degree of homogeneity in $\theta_\alpha$. The first field
$c$, the ghost field, has  zero degree of homogeneity in $\theta$
and negative internal parity. The second group of fields: scalar
$\varphi$, gauge field $A_\pm$ and boson fields $\gamma_\pm$ are
linear in $\theta$ and have positive internal parity. The fermion
fields $\psi_\pm$ and $\chi_\pm$ are quadratic in $\theta$ and have
again negative internal parity. Analyzing these examples one can
come to the conclusion that all the fields in the table have {\bf
negative total parity}, which is $\theta$-parity plus internal
parity. While all the antifields (the third column of the table)
have {\bf positive total parity}. This rule works in a universal way
for arbitrary system of quadrics $f^\mu(\lambda)$ and is the
peculiar feature of the formalism. One can check this rule for the
{\bf "regular"} system of section 5.3.

  The most non-trivial part of the effective action is the gauge sector.
By  straightforward calculation one can check that in this sector
the only fields contributing into in-lines are $\varphi$, $A_-$,
$A_+$ and the only out-line field is $\widetilde{\varphi}_8$. It
happens that only the field $\varphi$ contributes into the diagram
in the figure \ref{figS3}
which is responsible for the kinetic term in this sector. Applying
to this field operator $\Phi$
$$
\Phi(\lambda_2\theta_3 + \lambda_3\theta_2)\varphi\ =\
2\lambda_1\lambda_3\theta_1\theta_2\partial_+\varphi\ +\
2\lambda_2\lambda_4\theta_4\theta_3\partial_-\varphi
$$
then operator $K\ =\ Q^{-1}$
$$
K\ \!\Phi(\lambda_2\theta_3 + \lambda_3\theta_2)\varphi\ =\
2\lambda_1\theta_3\theta_1\theta_2\partial_+\varphi\ +\
2\lambda_4\theta_2\theta_4\theta_3\partial_-\varphi
$$
and finally operator $\Phi$ again
$$
\Phi\ \! K\ \!\Phi(\lambda_2\theta_3 + \lambda_3\theta_2)\varphi\ =\
4\lambda_1\lambda_4\theta_4\theta_3\theta_1\theta_2\partial_-\partial_+\varphi\
+\ 4\lambda_1\lambda_4\theta_1\theta_2\theta_4\theta_3
\partial_+\partial_-\varphi\Big|_{Projecting}\!\!\!\longrightarrow\
\ -8\widetilde{\varphi}_8\partial_+\partial_-\varphi
$$
results into the following contribution into effective action
\begin{equation}\label{S3}
S^{(3)}\ =\ \int\ STr \big(
-8\widetilde{\varphi}_8\partial_+\partial_-\varphi \big)
\end{equation}
The quartic terms are given by the  diagrams in figure \ref{figS4}

\vspace{-0.3cm} \noindent plus the same diagrams with interchanged
$A_+$ and $A_-$. All four contributions of figure \ref{figS4} are
different due to the clockwise rule of section 5.2. The calculation
of these diagrams is the same as that in figure \ref{4-point}. We
present \newpage  \noindent only the final answer which is a double
commutator of the fields:
\begin{equation}\label{S4}
S^{(4)}\ =\ \int STr \big(\
g^2\widetilde{\varphi}_8[A_-,[\varphi,A_+] ]\ +\
g^2\widetilde{\varphi}_8[A_+,[\varphi,A_-] ]\  \big)
\end{equation}

\noindent The last contribution is into cubic terms. Relevant
diagrams are presented in the figure \ref{figS5}.
\vspace{-0.3cm} \noindent In the first diagram possible combinations
of external in-lines are $(\varphi,A_-)$, $(A_-,\varphi)$,
$(\varphi,A_+)$, $(A_+,\varphi)$. Straightforward calculation of
these diagrams gives
\begin{equation}\label{S5}
S^{(5)}\ =\ \int\ STr \Big(
2g\widetilde{\varphi}_8[\partial_+\varphi, A_-]\ +\
2g\widetilde{\varphi}_8[\partial_-\varphi,A_+]\ +\
2g\widetilde{\varphi}_8\partial_-[\varphi,A_+]\ +\
2g\widetilde{\varphi}_8\partial_+[\varphi,A_-] \Big)
\end{equation}
These cubic diagrams are the last relevant contributions into
effective action for the gauge sector. Collecting together  all the
results of (\ref{S3}),(\ref{S4}),(\ref{S5}) one can write the
effective action for the gauge sector as:
\begin{equation}\label{Sgauge}
\begin{split}
S^{gauge}\ =\ \int\ STr\  \widetilde{\varphi}_8\Big(
-8\partial_+\partial_-\varphi\ +\ 2g[\partial_+\varphi,A_-]\ +\
2g[\partial_-\varphi,A_+]\ +\ 2g\partial_-[\varphi,A_+]\ +\ \\ +\
2g\partial_+[\varphi,A_-]\ +\ g^2[A_-,[\varphi,A_+]]\ +\
g^2[A_+,[\varphi,A_-]]\Big)
\end{split}
\end{equation}
It is possible to recognize in this expression in the brackets the
covariant laplacian in the adjoint representation
$-\partial_\mu\partial_\mu\varphi\ -\ g[A_\mu,\
\partial_\mu\varphi]\ -\ g\partial_\mu[A_\mu,\
\varphi]\ -\ g^2[A_\mu,\ [A_\mu,\ \varphi]]$
in two dimensions\footnote{Summation is with respect to the metric
$\eta_{\mu\nu}=diag(1,-1)$} under the identification
$$
{\begin{array}{cc} A_1=\frac{1}{\sqrt{2}}(A_+\ +\ A_-) \ \ \ \ & \ \
\ \
\partial_1\ =\
\sqrt{2}(\partial_+\ +\
\partial_-)\\
A_2=\frac{1}{\sqrt{2}}(A_+\ -\ A_-)\ \ \ \  &\ \ \ \  \partial_2\ =\
\sqrt{2}(\partial_+\ -\
\partial_-)
\end{array}}
$$

\noindent  At this step evaluation of effective action for the {\bf
"singular"} system of quadrics (\ref{singular constraints}) is
finished. As we explained in section 3 it is obtained from BV
integration over the lagrangian sub-manifold. Hence, the result is
again a BV action. The question is: "To what theory this BV action
corresponds?"

  Effective action for the system (\ref{singular constraints})
contains  totally 36 fields (18 fields and 18 antifields). Since
there is a complete symmetry between the fields and antifields in BV
formalism, for each pair field - antifield   one has a freedom to
choose any one of this pair and consider it as a field and another
one as an antifield. Then, to extract classical gauge invariant
action $I[A]$ from the BV action one can simply put all the
antifields equal to zero. This is obvious from the definition of BV
action (\ref{BVgauge}). For our particular model the convenient
choice is the following\footnote{Note, that imposing the condition -
all the antifields are equal to zero - determines the lagrangian
sub-manifold.}:
\begin{equation}
\begin{array}{|c|c|c|c|c|c|c|c|c|c|c|c|c|c|c|c|c|c|c|}
\hline
$ $  &  &  &  &  &  &  &  &  &  &  &  &  &  &  &  &  &  &  \\
$Fields$ &\  c \ & \gamma_+ &\  \varphi\  & \gamma_- & A_+ & A_- & \psi_+ & \psi_- & \chi_+ & \chi_- & \widetilde{\varphi}_1 & \widetilde{\varphi}_2 & \widetilde{\varphi}_3 & \widetilde{\varphi}_4 & \widetilde{\varphi}_5 & \widetilde{\varphi}_6 & \widetilde{\varphi}_7 & \widetilde{\varphi}_8 \\
$ $  &  &  &  &  &  &  &  &  &  &  &  &  &  &  &  &  &  &  \\
\hline
$ $  &  &  &  &  &  &  &  &  &  &  &  &  &  &  &  &  &  &  \\
$Antifields$  &\  \widetilde{c} \ & \widetilde{\gamma}_+ &\  \widetilde{\varphi}\  & \widetilde{\gamma}_- & \widetilde{A}_+ & \widetilde{A}_- & \widetilde{\psi}_+ & \widetilde{\psi}_- & \widetilde{\chi}_+ & \widetilde{\chi}_- & \varphi_1 & \varphi_2 & \varphi_3 & \varphi_4 & \varphi_5 & \varphi_6 & \varphi_7 & \varphi_8 \\
$ $  &  &  &  &  &  &  &  &  &  &  &  &  &  &  &  &  &  &  \\
\hline
\end{array}
\end{equation}
At the next step one should put all the antifields equal to zero.
Then the result for the classical action which we denote $I[A]$ is
given by:
\begin{equation}\label{classical action}
\begin{split}
I[A]\ =\ \int Tr\bigg( \ \ \widetilde{\varphi}_1\Big(\
2(\partial_+A_- -
\partial_-A_+)\ +\ g[A_+,\ A_-]\ \Big)\ + \widetilde{\varphi}_8\Big(
-8\partial_+\partial_-\varphi\ +\ 2g[\partial_+\varphi,A_-]\ +\
2g[\partial_-\varphi,A_+]\ +\\ +\  2g\partial_-[\varphi,A_+]\ +\
2g\partial_+[\varphi,A_-]\ +\ g^2[A_-,[\varphi,A_+]]\ +\
g^2[A_+,[\varphi,A_-]]\Big) -\
g\widetilde{\varphi}_8\{\psi_+,\ \psi_-\}\ +\\
 +\ \widetilde{\varphi}_2\Big(\ \ 2\partial_-\gamma_+\ +\ g[A_-,
\gamma_+]\ \Big)\ +\ \widetilde{\varphi}_3\Big(\
2\partial_+\gamma_-\ +\ g[A_+,\gamma_-]\ \Big)\ +\
\widetilde{\varphi}_4\Big(\ 2\partial_+\psi_-\ +\ g[A_+, \psi_-]\
\Big)\ +\\ +\  \widetilde{\varphi}_5\Big(\ 2\partial_-\psi_+\ +\
g[A_-, \psi_+]\ \Big)\ +\ \widetilde{\varphi}_6\Big(\
\partial_+\chi_-\ +\ \frac{1}{2}g[A_+, \chi_-]\ +\
g[\gamma_-,\psi_+]\ \Big)\ +\\ +\ \widetilde{\varphi}_7\Big(\
\partial_-\chi_+\ +\ \frac{1}{2}g[A_-, \chi_+]\ +\
g[\gamma_+,\psi_-]\ \Big)\ \bigg)
\end{split}
\end{equation}
Above we have mentioned that the total parity ($\theta$ parity plus
internal parity) is always negative for the AKSZ field. Since
internal parity for AKSZ antifield is negative, the total parity of
AKSZ antifield (the fields in the third column of the table on the
page 22) is positive. Hence, in the action $I[A]$ internal parities
are given by:
$$
\begin{array}{|c|c|c|c|c|c|c|c|c|c|c|c|c|c|c|c|c|c|}
\hline
 &  &  &  &  &  &  &  &  &    \\
$Positive internal parity$  & \gamma_+ & \varphi & \gamma_- & A_+ & A_- & \widetilde{\varphi}_1 & \widetilde{\varphi}_2 & \widetilde{\varphi}_3 & \widetilde{\varphi}_8    \\
 &  &  &  &  &  &  &  &  &    \\
\hline
 &  &  &  &  &  &  &  &  &    \\
$Negative internal parity$  & \psi_+ & \psi_- & \chi_+ & \chi_- & \widetilde{\varphi}_4 & \widetilde{\varphi}_5 & \widetilde{\varphi}_6 & \widetilde{\varphi}_7 &      \\
 &  &  &  &  &  &  &  &  &    \\
\hline
\end{array}
$$
Thus, we identify the fields with positive internal parity with
bosons and with negative internal parity with fermions. Hence,
introducing notations for the fermions $\overline{\psi}$,
$\overline{\chi}$ and bosons  $\beta$, $\Phi$, $\phi_1$, $\phi_2$ in
the following way:
$$
\left\{\begin{array}{c} \widetilde{\varphi}_4\ =\ \overline{\psi}_-\\
\widetilde{\varphi}_5\ =\ \overline{\psi}_+
\end{array}\right.\ \ \ \ \ \ \ \ \ \left\{\begin{array}{c} \widetilde{\varphi}_6\ =\ 2\overline{\chi}_-\\
\widetilde{\varphi}_7\ =\ 2\overline{\chi}_+
\end{array}\right.\ \ \ \ \ \ \ \ \ \left\{\begin{array}{c} \widetilde{\varphi}_2\ =\ \beta_+\\
\widetilde{\varphi}_3\ =\ \beta_-
\end{array}\right.\ \ \ \ \ \ \widetilde{\varphi}_1\ =\ \Phi\ \ \ \
\ \ \ \ \left\{\begin{array}{c}\!\!\! \varphi\ =\ \frac{1}{\sqrt{2}}(\phi_1\ +\ i\phi_2)\\
\widetilde{\varphi}_8\ =\ \frac{1}{\sqrt{2}}(\phi_1\ -\ i\phi_2)
\end{array}\right.
$$
one can write the action (\ref{classical action}) in the canonical
form:
\begin{equation}\label{final answer for the effective action}
\begin{split}
I[A]\ =\ \int\ Tr\ \bigg(\ \Phi F_{+ -}\ +\ D_+\phi_1D_-\phi_1\ +\
D_-\phi_2D_+\phi_2\ -\ \frac{g}{\sqrt{2}}\phi_1\{\psi_+,\psi_-\}\ +\
i\frac{g}{\sqrt{2}}\phi_2\{\psi_+,\psi_-\}\ +\ \beta_+D_-\gamma_+\
+\\ +\ \beta_-D_+\gamma_-\ +\ \overline{\psi}_-D_+\psi_-\ +\
\overline{\psi}_+D_-\psi_+\ +\ \overline{\chi}_-D_+\chi_-\ +\
\overline{\chi}_+D_-\chi_+\ +\ 2g\overline{\chi}_-[\gamma_-,\psi_+]\
+\ 2g\overline{\chi}_+[\gamma_+,\psi_-]\ \bigg)
\end{split}
\end{equation}
It should be emphasized that the fields $\overline{\psi}$ and
$\overline{\chi}$ are not connected with $\psi$ and $\chi$ by
complex conjugation. They are other dynamical fields of the theory.
We also use notations\ \  $D_+\ =\ 2\partial_+\ +\ g[A_+, \ \ ]$\ \
and\ \ $D_-\ =\ 2\partial_-\ +\ g[A_-, \ \ ]$\ \ for left and right
adjoint covariant derivatives and $F_{-+}\ =\ 2(\partial_-A_+\ -\
\partial_+A_-)\ +\ g[A_-,A_+]\ =$
$=\ \partial_1A_2\ -\ \partial_2A_1\ +\ g[A_1,A_2]\ =\ F_{12}$ for
the field strength.

  From the expression  (\ref{final answer for the effective action})
it is clear that the spectrum of  effective theory contains: a gauge
field (which is not dynamical), two adjoint scalars $\phi_1$ and
$\phi_2$, fermions $\psi$, $\overline{\psi}$, $\chi$,
$\overline{\chi}$ and bosonic fields $\beta$ and $\gamma$ with the
kinetic term of first order in derivatives. The field $\Phi$ plays
the role of Lagrange multiplier for the zero curvature condition.
\\

 The theory (\ref{final answer for the effective action}) provides
another example of physically interesting theory which arises from
Berkovits construction or equivalently from the action (\ref{Fund}).
Though the gauge field is not dynamical in this theory, there is a
non-trivial Yukawa-like interaction between left and right fermions.
There are also scalars which have kinetic term quadratic in
derivatives. We have already mentioned in the introduction that this
theory is the minimal model, complying with this condition (minimal
number $K=4$ of $\theta_\alpha$ and minimal number $N=5$ of quadrics
$f^\mu(\lambda)$\ ). Thus, it is of interest to search for other
examples of {\bf "singular"} quadrics which provide non-trivial
physically interesting theories arising from the fundamental action
(\ref{Fund}). Such theories will reproduce non-trivial gauge models
in $d=3$ and probably even in $d=4$ dimensions.

\newpage

\section{Conclusion}
 It was shown that the BV action for a large class of field theories
can be obtained as effective action from the fundamental theory
(\ref{Fund}). The whole information about degrees of freedom and the
structure of effective action is encoded in the system of quadrics
$f^\mu(\lambda)$. There are two classes of such quadrics: {\bf
"regular"} and {\bf "singular"}. Effective action for {\bf
"regular"} system is universal, independently of the particular
$\lambda_\alpha$ dependence of $f^\mu(\lambda)$. Effective action
for {\bf "singular"} quadrics is not universal and strongly depends
on the structure of $f^\mu(\lambda)$. One example of such {\bf
"singular"} theories is the 2-D model of section 6.

\begin{figure}[h]
\centerline{\includegraphics[width=70mm,angle=-90]{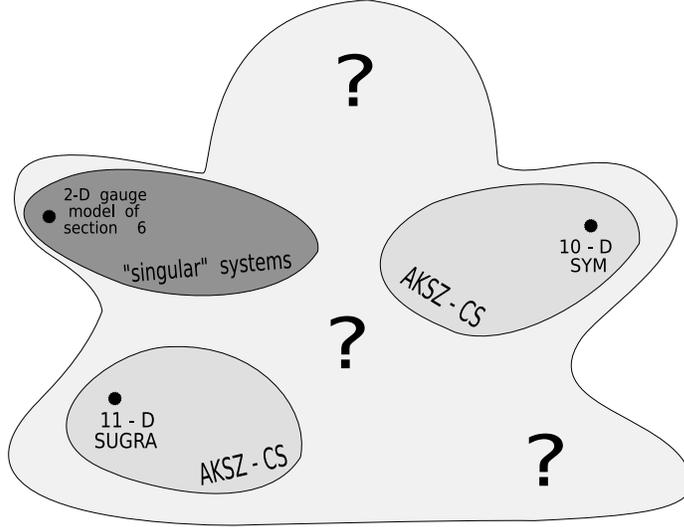}}
\caption{{\footnotesize An illustration on the structure of the
space of effective theories.}} \label{conclusion}
\end{figure}

As we mentioned in the introduction these ideas can be depicted in
the picture (see figure \ref{landscape}) illustrating the structure
of the space of effective theories. As the last concluding remark we
would like to clarify this picture a little (see figure
\ref{conclusion}). \vspace{-0.5cm}

\noindent Different {\bf "singular"} theories (black points)
representing 10-D SYM, 11-D SUGRA, 2-D model of section 6 have
different numbers $K$ of $\lambda_\alpha$ and $N$ - the number of
quadrics. Hence they are from the different domains of the space of
effective theories (see figure \ref{conclusion}). The point is that
in case $K\ \geq\ N$ there exist both {\bf "regular"} and {\bf
"singular"} systems of quadrics. However if $N\ >\ K$ there are no
{\bf "regular"} systems. All quadrics are {\bf "singular"} in this
case. Hence the theories 10-D SYM ($K=16$, $N=10$) and 11-D SUGRA
($K=32$, $N=11$) are indeed surrounded by AKSZ-CS theories in their
domains (strictly speaking the theories 10-D SYM and 11-D SUGRA are
obtained from the effective action only after $Z_2$ reduction, which
is similar to the one discussed in section 2). However the 2-D model
of section 6 ($K\ =\ 4$, $N\ =\ 5$) is surrounded again by singular
theories. This fact is indicated in the figure \ref{conclusion}.
Even in case $K\geq N$ there could be small deformations of quadrics
which remain the system $f^\mu(\lambda)$ {\bf "singular"}. Hence,
even in this case {\bf "singular"} theories are not single dots but
can have some modules. The signs "?" stand in the figure because we
do not know whether it is possible or not to deform the construction
continuously from the one domain to an another.

\vspace{-0.3cm}
\section{Appendix. Algebraic Meaning of the Berkovits
Complex}


\bigskip
\centerline{ABSTRACT}

\bigskip
   Complete "mathematical" proof of the theorem from the section 4 is given.
Using the Zig-Zag (Tick-Tack-Toe) technique it is demonstrated that
cohomologies of the Berkovits complex are in one-to-one
correspondence with the cohomologies of certain complex with zero
differential. This allows at once to write down explicit expressions
(\ref{coh}) for all representatives of $\mathcal{H}(Q,O)$. \\
\centerline{\underline{\hspace{5.5cm}}}

\bigskip
\noindent In the main part of the paper we have studied the complex
$$
C[\lambda,\theta]/I_f
$$
with the differential
$$
Q=\lambda^{\alpha}\frac{\p}{\p \theta^{\alpha}}
$$
where $I_f$ is the ideal in $C[\lambda]$ generated by a set of
quadrics $f_{\mu}(\lambda)$. We are interested in this complex
because its cohomologies correspond to polarizations of fields in
various (supersymmetric) quantum field theories.

At first glance this complex looks like another trick in
supersymmetry. However (like many other tricks) this one is also
related to some algebro-geometric construction. One can show  that
Berkovits complex is nothing but the complex computing the higher
derived functor $ Tor^{*} $ between two modules over the ring
$C[\lambda]$.

Here we will give an elementary explanation of this fact. It would
also explain why cohomologies of Berkovits complex are computed in
terms of relations between relations.

\begin{figure}[h]
\centerline{\includegraphics[width=160mm]{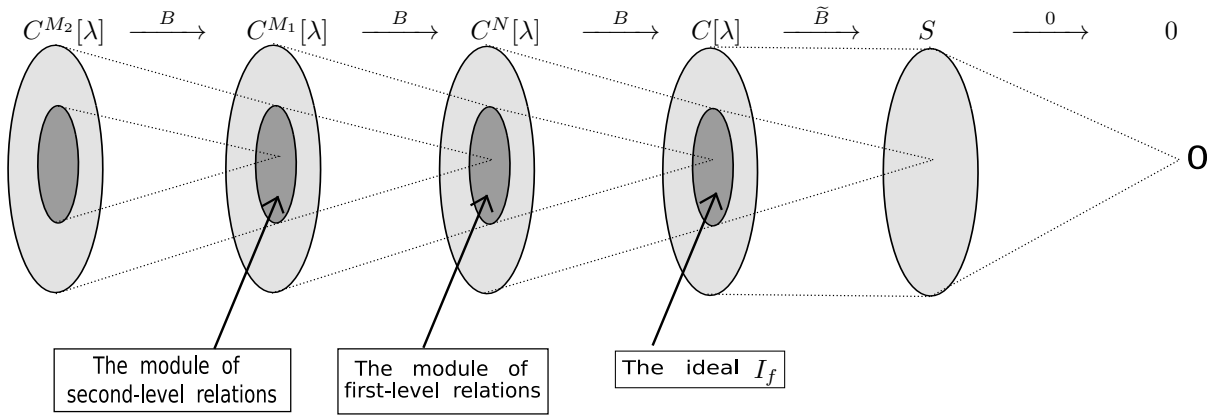}} \caption[
]{\label{free resolvent}  {\footnotesize The free
resolution.}}\noindent
\end{figure}

First, let us explain what is the {\bf minimal free resolution}.
Consider the sequence:
\begin{equation}\label{CD for free resolvent}
\begin{CD}
.......\ @>B>>\ C^{ M_2}[\lambda]\ @>B>>\ C^{ M_1}[\lambda]\ @>B>>\
C^{ N}[\lambda]\ @>B>>\ C[\lambda]\ @>\widetilde{B}>>\ S\ @>0>>\ 0
\end{CD}
\end{equation}
Here $C[\lambda]$ is a ring of polynomials, $C^{ N}[\lambda]$, $C^{
M_1}[\lambda]$, $C^{ M_2}[\lambda]$ is respectively $N$, $M_1$ and
$M_2$ copies of $C[\lambda]$, and $S$ is the coset
$S=C[\lambda]/I_f$. The differential in this complex is defined as
follows: the first differential from the right is simply zero, the
second one - $\widetilde{B}$ makes factorization over the ideal
$I_f$. To define other maps introduce a basis in the spaces $C^{
N}[\lambda]$, $C^{ M_1}[\lambda]$, $C^{ M_2}[\lambda]$, ... The
basis elements are $e_\mu$ ($\mu=1..N$), $e_{A_1}$ ($A_1=1..M_1$),
$e_{A_2}$ ($A_2=1..M_2$),... The action of $B$ is given by:
\begin{equation}\label{basis for free resolvent}
\begin{array}{c}
B(e_\mu)\ =\ f_\mu\\
B(e_{A_1})\ =\ G_{A_1}^{(1)\ \mu}\  e_\mu\\
B(e_{A_2})\ =\ G_{A_2}^{(2)\ A_1}\  e_{A_1}\\
..............................................
\end{array}
\end{equation}
where $G_{A_1}^{(1)\ \mu}(\lambda)$ is the fundamental system of
relations at the first level, $G_{A_2}^{(2)\ A_1}(\lambda)$ are the
second-level fundamental relations and so on. This sequence is
schematically depicted in the figure \ref{free resolvent}.
$$
\begin{CD}
\!\!\! C^{ M_2}[\lambda]\ @>B>>\ \ \ C^{ M_1}[\lambda]\ \ @>B>>\ \
C^{ N}[\lambda]\ \ \ @>B>>\ \ C[\lambda]\ \ \ @>\widetilde{B}>>\ \ \
\ \ S\ \ \ \ \ \ \ @>0>>\ \ \ \ \ \ \ 0
\end{CD}
$$
\vspace{-1.1cm}
\begin{picture}(10,10)(30,30)
\put(324,-77){$I_f$}
\end{picture}

In this figure light ovals denote the spaces $S$, $C[\lambda]$, $C^{
N}[\lambda]$, $C^{ M_1}[\lambda]$,... Each space contains a subspace
(dark oval) which is mapped to zero by the differential. For
example, operator $\widetilde{B}:\ C[\lambda]\ \longrightarrow\ S$
maps each element of the ideal $I_f$ (dark oval in $C[\lambda]$) to
zero. At the previous step differential $B:\ C^{ N}[\lambda]\
\longrightarrow\ C[\lambda]$ maps an arbitrary element
$P^\mu(\lambda)e_\mu$ of $C^{ N}[\lambda]$ onto
$B(P^\mu(\lambda)e_\mu)\ =\ P^\mu(\lambda)f_\mu(\lambda)$ - the
element of the ideal. Hence, the pre-image of the ideal $I_f$ is the
whole space $C^{ N}[\lambda]$. This fact is illustrated in the
figure \ref{free resolvent} by the conic-like dashed lines. The free
module $C^{ N}[\lambda]$ again contains the subspace which is mapped
to zero. This subspace is given by the elements
$P^\mu(\lambda)e_\mu$, satisfying $B(P^\mu(\lambda)e_\mu)\ =\
P^\mu(\lambda)f_\mu(\lambda)\ =\ 0$. Hence, the dark oval in $C^{
N}[\lambda]$ is the module of first-level relations. Again, the
pre-image of this module is the whole space $C^{ M_1}[\lambda]$.
Indeed, the arbitrary element $P^A(\lambda)e_A$ is mapped to
$B(P^A(\lambda)e_A)\ =\ P^A(\lambda)G_A^{(1)\ \mu} e_\mu$ which is a
relation since $G_A^{(1)\ \mu}(\lambda)$ is a fundamental relation
(\ $G_A^{(1)\ \mu}f_\mu(\lambda)\ =\ 0$\ ). Similar arguments can be
applied to higher terms $C^{ M_2}[\lambda]$,\ \ $C^{ M_3}[\lambda]$
and so on. From the figure \ref{free resolvent}
it is obvious that the sequence satisfies two properties:\\
\begin{enumerate}
\item Double action of the differential on an arbitrary element
of the sequence gives zero (this is clear from the dashed lines
in figure \ref{free resolvent}):\ $B^2\ =\ 0$.\\
\item This sequence has no cohomologies: each closed element is
exact (again, this follows from the dashed lines).\\
\end{enumerate}
The sequence, satisfying these two conditions is called the {\bf
minimal free resolution}. Minimality means that the elements of the
fundamental system of relations are linearly independent. In this
sequence most of terms are n-copies of the ring $C[\lambda]$. Hence,
they are free modules. However, the first term $S$ is given by the
coset, hence is far from being free.

  Now we are ready to present the crucial instrument of the
construction. Consider the bi-complex with differentials $B$ and $Q$
which can be written as the table:
\begin{equation}\label{main table}
\begin{CD}
.....@>Q>>\ [F(\lambda,\theta^2)]\ @>Q>>\ [F(\lambda,\theta)]\ @>Q>>\ [F(\lambda)]\\
@.  @AA\widetilde{B}A @AA\widetilde{B}A @AA\widetilde{B}A @.\\
.....@>Q>>\ F(\lambda,\theta^2)\ @>Q>>\ F(\lambda,\theta)\ @>Q>>\ F(\lambda)\ @>\widetilde{Q}>>\ c\\
@.  @AABA  @AABA @AABA @AA0A @.\\
.....@>Q>>\ F^\mu(\lambda,\theta^2)e_\mu\ @>Q>>\ F^\mu(\lambda,\theta)e_\mu\ @>Q>>\ F^\mu(\lambda)e_\mu\ @>\widetilde{Q}>>\ c^\mu e_\mu\\
@.  @AABA  @AABA @AABA @AA0A @.\\
.....@>Q>>\ F^{A_1}(\lambda,\theta^2)e_{A_1}\ @>Q>>\ F^{A_1}(\lambda,\theta)e_{A_1}\ @>Q>>\ F^{A_1}(\lambda)e_{A_1}\ @>\widetilde{Q}>>\ c^Ae_A \\
@.  @AABA  @AABA @AABA @AA0A @.\\
@ ...... @ ...... @ ...... @ ......\\
\end{CD}
\end{equation}
\begin{picture}(10,10)(30,30)
\Line(397,45)(397,193)\Line(50,193)(397,193)
\end{picture}
The first line - the horizontal margin of the table - is the
Berkovits complex. The differential is given by
$Q=\lambda_\alpha\frac{\partial}{\partial\theta_\alpha}$ which acts
in the space $O[\theta_\alpha\ \! ,\ \! \lambda_\alpha\ |\
f^{\mu}(\lambda)]$. The grading in this space is given by the degree
of $\theta$ and $Q$ decreases the grading. We use the notations
$[F(\lambda)]$,\ \ $[F(\lambda,\theta)]$,\ \
$[F(\lambda,\theta^2)]$,... to represent equivalence classes of
functions, of certain degree in $\theta_\alpha$, in the space
$O[\theta_\alpha\ \! ,\ \! \lambda_\alpha\ |\ f^{\mu}(\lambda)]\ =\
C[\lambda,\theta]\diagup I_f$.

Each term of the Berkovits complex is resolved in vertical direction
by the free resolution of figure \ref{free resolvent}. This means
that one can think about each column in the table as the tensor
product the free resolution by the functions of certain degree in
$\theta_\alpha$. This resolution is equipped with the action of the
differential $B$, defined in (\ref{basis for free resolvent}), and
$\widetilde{B}$ which acts when one jumps from the table to the
horizontal margin of the table (the sequence above the horizontal
black line)\footnote{We emphasize here that the notation
$F(\lambda,\theta^2)$ below the black line is used to denote a
function, quadratic in $\theta_\alpha$. Notation
$[F(\lambda,\theta^2)]$  is used to denote an equivalence class
which is obtained from $F(\lambda,\theta^2)$ after factorization
over the ideal $I_f$.}.

Thus we see that the table is equipped with the structure of the
free resolution in the vertical direction. It happens that it is
possible to build the complex on the vertical margin of the table
(to the right of the black vertical line) in such a way that each
horizontal sequence will be again a free resolution. To demonstrate
this take as an example the second line of the table:
\begin{equation}\label{horizontal free resolvent}
\begin{CD}
.....@>Q>>\ F^\mu(\lambda,\theta^2)e_\mu\ @>Q>>\
F^\mu(\lambda,\theta)e_\mu\ @>Q>>\ F^\mu(\lambda)e_\mu\
@>\widetilde{Q}>>\ c^\mu e_\mu\ @>0>>\ 0
\end{CD}
\end{equation}
Since within the table (below the black horizontal line and to the
left from the vertical one) there are no any constraints, like
$f^\mu(\lambda)$ , the sequence (\ref{horizontal free resolvent}) is
free of cohomologies in the sector of $\theta$, $\theta^2$,
$\theta^3$ and so on. This is true due to $Ker (Q)\ =\ 0$ in the
ring $C[\lambda,\theta]$ in these sectors. Still, the question
remains in the sector $F^\mu(\lambda)e_\mu$ of zero degree in
$\theta$. To make the sequence exact, we define the
$\widetilde{Q}$-operator (the jump from the table to the vertical
margin) as factorization over the ideal $I_\lambda$, generated by
the elements $\{\lambda_1,\lambda_2,...,\lambda_{K}\}$. In this case
the image of $Q$-operator are functions
$Q(F^\mu(\lambda,\theta)e_\mu)$ at least linear in $\lambda_\alpha$
( since $Q\ =\ \lambda_\alpha\frac{\partial}{\partial\theta_\alpha}$
). Hence, operator $\widetilde{Q}$ (factorization over the ideal
$I_\lambda$) maps such functions to zero, removing all cohomologies,
in the sector $F^\mu(\lambda)e_\mu$. Thus to make the sequence
(\ref{horizontal free resolvent}) exact the complex on the vertical
margin should be defined in the space $C[\lambda]\diagup I_\lambda\
=\ \mathbb{C}$ of complex numbers. Hence, the differential $B$,
which is defined through relations (\ref{basis for free resolvent})
is simply zero on the vertical margin, due to the fact that
fundamental relations $G^{(1)}(\lambda)$, $G^{(2)}(\lambda)$,... are
at least linear in $\lambda_\alpha$. This fact is indicated on the
vertical margin of (\ref{main table}) where the differential $B$ is
reduced to $B\ =\ 0$ because of factorization over the ideal
$I_\lambda$.

  Concluding the presentation of the bi-complex, we summarize that
it is written as the table (below the black horizontal line and to
the left from the vertical one). This table has two margins:
horizontal (above the horizontal line) and vertical (to the right of
the vertical line). Horizontal margin is the Berkovits complex with
operator $Q$, acting in the space $O[\theta_\alpha\ \! ,\ \!
\lambda_\alpha\ |\ f^{\mu}(\lambda)]$. Vertical margin is the
complex with zero differential, acting in the spaces
$C[\lambda]\diagup I_\lambda\ =\ c$,$ $ $\ C^{ N}[\lambda]\diagup
I_\lambda\ =\ c^\mu e_\mu$, $ $ $\ C^{ M_1}[\lambda]\diagup
I_\lambda\ =\ c^{A_1} e_{A_1}$ and so on. Here $c$, $c^\mu$,
$c^{A_1}$, ..... are simply complex numbers. All the terms within
the table are free of any factorization conditions.

  Our last remark is that the action of  operators $B$ and $Q$
is commutative $[B,\ Q]\ =\ 0$ on the table.  This is true since the
action of $B$ is defined through the fundamental system of relations
(\ref{basis for free resolvent}), hence contains only multiplication
by $\lambda$. Operator $Q$ also increases the number of $\lambda$.
Since $\lambda_\alpha$ are commuting variables $[B,\ Q]\ =\ 0$. For
the same reason commutators $[B,\ \widetilde{Q}]\ =\ 0$ and
$[\widetilde{B},\ Q]\ =\ 0$ also vanish.

\bigskip

\noindent
\underline{{\bf The Zig-Zag Theorem:}} the cohomologies of the
Berkovits complex $\mathcal{H}(Q,O)$ are in one-to-one
correspondence with the cohomologies of the complex at the vertical
margin of the table.

\bigskip

\noindent The proof is in zig-zag jumps on the table (\ref{main
table}). All arguments are similar for each term of the Berkovits
complex. Thus we illustrate them taking the first non-trivial
example - the class $[F(\lambda,\theta)]$. The zig-zag responsible
for this term is presented in the table:
\begin{equation}\label{first zig-zag table}
\begin{CD}
.....@>Q>>\ [F(\lambda,\theta^2)]\ @>Q>>\ [F(\lambda,\theta)]\ @>Q>>\ [F(\lambda)]\\
@.  @AA\widetilde{B}A @AA\widetilde{B}A @AA\widetilde{B}A @.\\
.....@>Q>>\ F(\lambda,\theta^2)\ @>Q>>\ F(\lambda,\theta)\ @>Q>>\ F(\lambda)\ @>\widetilde{Q}>>\ c\\
@.  @AABA  @AABA @AABA @AA0A @.\\
.....@>Q>>\ F^\mu(\lambda,\theta^2)e_\mu\ @>Q>>\ F^\mu(\lambda,\theta)e_\mu\ @>Q>>\ F^\mu(\lambda)e_\mu\ @>\widetilde{Q}>>\ c^\mu e_\mu\\
@.  @AABA  @AABA @AABA @AA0A @.\\
.....@>Q>>\ F^{A_1}(\lambda,\theta^2)e_{A_1}\ @>Q>>\ F^{A_1}(\lambda,\theta)e_{A_1}\ @>Q>>\ F^{A_1}(\lambda)e_{A_1}\ @>\widetilde{Q}>>\ c^Ae_A \\
@.  @AABA  @AABA @AABA @AA0A @.\\
@ ...... @ ...... @ ...... @ ......\\
\end{CD}
\end{equation}
\begin{picture}(10,10)(30,30)
\Line(387,150)(387,193)\Line(300,193)(387,193)
\Line(387,150)(460,150)\Line(300,193)(300,230)
\end{picture}

In the {\it\underline{step1}} of the proof we show that each
$Q$-closed element in the space $[F(\lambda,\theta)]$ has a
corresponding element in the sector $c^\mu e_\mu$ of the vertical
complex. Since the complex has zero differential this element is
$0$-closed. In the {\it\underline{step2}} we show that each
$Q$-exact element in the sector $[F(\lambda,\theta)]$ is mapped to
zero in the sector $c^\mu e_\mu$ (by the same zig-zag). Since the
vertical complex has zero differential, it has no exact expressions.
Hence, according to {\it\underline{step1}} and
{\it\underline{step2}} we conclude that there is a map from the
cohomologies of the Berkovits complex to the cohomologies of the
vertical complex. The only question remains if this map is a
one-to-one correspondence. To convince that it is the case, in
{\it\underline{step3}} we start from the certain element in the
sector $[F(\lambda,\theta)]$, then pass through the zig-zag to the
vertical complex and then, along the same zig-zag, in the opposite
direction. The fact that we obtain the same element completes the
proof  that the zig-zag map is the one-to-one correspondence.

\bigskip

\noindent{\it\underline{step1}}\ \ \ \  Suppose the class
$[F(\lambda,\theta)]$ is $Q$-closed. This means that $Q
F(\lambda,\theta)\ =\ c^\mu(\lambda)f_\mu(\lambda)$, where
$c^\mu(\lambda)$ are some coefficients. We are going to pass along
the zig-zag, indicated in (\ref{first zig-zag table}), to the
vertical complex and find the corresponding element $c^\mu\
\!\!e_\mu$ in it. This element is $0$-closed.  Since the vertical
column
$$\begin{CD}..... @>B>>\ F^\mu(\lambda,\theta)e_\mu\ @>B>>\
F(\lambda,\theta)\ @>\widetilde{B}>>\ [F(\lambda,\theta)]\ @>0>>\ 0
\end{CD}$$ is the free resolution of figure \ref{free
resolvent} and the last term $[F(\lambda,\theta)]$ in it is mapped
to zero,  one can apply operator $\widetilde{B}^{-1}$ to
$[F(\lambda,\theta)]$, because the free resolution is free from
cohomologies.  Define the action of $\widetilde{B}^{-1}$ as
$$
\widetilde{B}^{-1}\Big([F(\lambda,\theta)]\Big)\ =\
F(\lambda,\theta)\
$$
Now we have passed through the first step of the zig-zag and have
jumped from the horizontal margin into the space $F(\lambda,\theta)$
of the table. Next step of the zig-zag is application of
$Q$-operator. The result is:
\begin{equation}\label{one step of zig-zag}
Q\Big(\widetilde{B}^{-1}[F(\lambda,\theta)]\ \ +\ \
\widetilde{B}-\!\!\!\begin{array}{c}$closed$\end{array} \Big)
\end{equation}
Applying $\widetilde{B}^{-1}$ we have a freedom to add
$\widetilde{B}$-closed expression. Now we have jumped into the space
$F(\lambda)$ on the table (\ref{first zig-zag table}). Next step is
application of $B^{-1}$. Since the vertical sequence is exact this
is possible only if $\widetilde{B}
Q\Big(\widetilde{B}^{-1}[F(\lambda,\theta)]\ +\ \
\widetilde{B}\!-\!\!\!\begin{array}{c}$closed$\end{array}\Big)\ =\
0$. Simple computation gives: \vspace{-0.2cm}
$$
\widetilde{B}Q\Big(\widetilde{B}^{-1}[F(\lambda,\theta)]\ +\ \
\widetilde{B}\!-\!\!\!\begin{array}{c}$closed$\end{array}\Big)\ =\ Q
\widetilde{B}\Big(\widetilde{B}^{-1}[F(\lambda,\theta)]\ +\ \
\widetilde{B}\!-\!\!\!\begin{array}{c}$closed$\end{array}\Big)\ =\
Q\widetilde{B}\widetilde{B}^{-1}[F(\lambda,\theta)]\ =\
Q[F(\lambda,\theta)]\ =\ 0
$$
In the first equality we used that $[Q,\ \widetilde{B}]\ =\ 0$, in
the second one that $\widetilde{B}(\
\widetilde{B}\!-\!\!\!\begin{array}{c}$closed$\end{array})\ =\ 0$.
In the last equality we explore that the class $[F(\lambda,\theta)]$
is $Q$-closed. Thus, it is possible to apply operator $B^{-1}$ to
(\ref{one step of zig-zag}). Now we have come to the term
$F^\mu(\lambda)e_\mu$ in the table (\ref{first zig-zag table}). The
last step of the zig-zag is to jump from the table to the vertical
margin. This can be done by application of operator $\widetilde{Q}$.
The result is:
$$
\widetilde{Q}\bigg(B^{-1}\ Q
\Big(\widetilde{B}^{-1}[F(\lambda,\theta)]\ +\
\widetilde{B}\!-\!\!\!\begin{array}{c}$closed$\end{array} \Big)\ +\
B\!-\!\!\!\begin{array}{c}$closed$\end{array}\bigg)
$$
This expression is a representative of the class $c^\mu e_\mu$ of
the complex on the vertical margin. Since the differential in this
vertical complex is zero, this expression realizes the map from
$Q$-closed classes $[F(\lambda,\theta)]$ to the $0$-closed elements
of the vertical complex. It should be emphasized  that this zig-zag
can be done only in case $[F(\lambda,\theta)]$ is $Q$-closed.

  Above we have done detailed calculation, emphasizing all
ambiguities  which can arise while inverting operator $B$. However
it is clear, that since we used explicit basis for defining the
action of $B$, it is possible to define $B^{-1}$ as
\begin{equation}\label{basis for inverse operator}
\begin{array}{c}
B^{-1}(f_\mu(\lambda))\ =\ e_\mu\\
B^{-1}(G_A^{(1)\ \mu} e_\mu)\ =\ e_A\\
..........................................
\end{array}
\end{equation}
The same is true for the operator $Q^{-1}$. According to (\ref{deg})
anticommutator  $\{D,\ Q\}\ =\ deg$, hence one can think that
$Q^{-1}\ =\ \widetilde{D}\ =\ \frac{D}{deg}$. For the following we
use these explicit expressions for the inverse operators, however
this is not necessary for completing the proof.

\bigskip

\noindent{\it\underline{step2}}\ \ \ \ Suppose we start from exact
class $[F(\lambda,\theta)]\ =\ Q[G(\lambda,\theta^2)]$. This means
that $[F(\lambda,\theta)]\ =\ Q G(\lambda,\theta)\ +\
c^\mu(\lambda,\theta)f_\mu(\lambda)$. Application of
$\widetilde{B}^{-1}$ to such class gives:
$$
\widetilde{B}^{-1}[F(\lambda,\theta)]\ =\ Q G(\lambda,\theta)\ +\
c^\mu(\lambda,\theta)f_\mu(\lambda)
$$
with probably another coefficients $c^\mu(\lambda,\theta)$. Applying
operator $Q$ to this expression one can get
$$
Q\widetilde{B}^{-1}[F(\lambda,\theta)]\ =\
Q\big(c^\mu(\lambda,\theta)\big)\ f_\mu(\lambda)
$$
Application of $\widetilde{B}$ to this result gives zero, since
$Q(c^\mu(\lambda,\theta))\ f_\mu(\lambda)$ belongs to the ideal
$I_f$. Hence, one can apply $B^{-1}$:
\begin{equation}\label{another step of zig-zag}
B^{-1}Q\widetilde{B}^{-1}[F(\lambda,\theta)]\ =\
Q\big(c^\mu(\lambda,\theta)\big)\ e_\mu
\end{equation}
The last step is application of $\widetilde{Q}$ to this result.
Since the function $c^\mu(\lambda,\theta)$ is linear in
$\theta_\alpha$, expression $Q c^\mu(\lambda,\theta)$ is at least
linear in $\lambda_\alpha$, because $Q\ =\
\lambda_\alpha\frac{\partial}{\partial\theta_\alpha}$. Hence the
application of $\widetilde{Q}$ to (\ref{another step of zig-zag})
gives zero, since $\widetilde{Q}$ is factorization over the ideal
$I_\lambda$. This completes the proof that all $Q$-exact classes on
the horizontal margin of (\ref{first zig-zag table}) are mapped to
zero on the vertical margin.

\bigskip

\noindent{\it\underline{step3}}\ \ \ \ After fixation of all
ambiguities in the inverting of operators $Q$ and $B$,
{\it\underline{step3}} is almost obvious. Suppose $A$ is the element
of cohomologies $\mathcal{H}(Q,\ O)$ in the sector
$[F(\lambda,\theta)]$. Following the zig-zag, we come to the
vertical margin by applying
$$
\widetilde{Q}\ B^{-1}\ Q\ \widetilde{B}^{-1}\ A
$$
Passing through the same zig-zag in the opposite direction one
obtains
$$
\widetilde{B}\ Q^{-1}\ B\ \widetilde{Q}^{-1}\ \widetilde{Q}\ B^{-1}\
Q\ \widetilde{B}^{-1}\ A\ \ =\ \ A
$$
One should also take into account the possibility to pass from an
arbitrary element $c^\mu e_\mu$ to the corresponding element in the
sector $[F(\lambda,\theta)]$. The arguments are similar to that of
{\it\underline{step1}} and {\it\underline{step2}}. First one obtains
$\widetilde{Q}^{-1}\big(c^\mu e_\mu\big)\ =\ c^\mu e_\mu$ - the
element of $F^\mu(\lambda)e_\mu$. Then, applying operator $B$, we
get the element $B\big(c^\mu e_\mu\big)\ =\ c^\mu f_\mu(\lambda)$ of
$F(\lambda)$. Since $f^\mu(\lambda)$ are quadratic in
$\lambda_\alpha$, action of operator $\widetilde{Q}(c^\mu f_\mu)\ =\
0$ vanishes. Hence, one can apply operator $Q^{-1}$, because the
horizontal sequence is exact. After that we have jumped into the
space $F(\lambda,\theta)$. Last step is application of operator
$\widetilde{B}$.

\noindent The same construction can be straightforwardly realized
for higher degrees in $\theta_\alpha$. Hence the proof of
{\bf The Zig-Zag Theorem} is finished.\\
\centerline{\underline{\hspace{5.5cm}}} \vspace{0.1cm}

Our last remark is that calculation of Berkovits cohomologies on the
horizontal margin of the table is a difficult mathematical problem.
However, calculation of cohomologies on the vertical margin is
almost automatic. Since differential in this complex is zero, the
whole space of complex coincides with its cohomologies. Hence, to
calculate the cohomology of Berkovits complex one can take a space
from the vertical complex, for example $c^\mu\ e_\mu$ (where $c^\mu$
are arbitrary numerical coefficients) and apply zig-zag to it.
\begin{equation}\label{first zig-zag formula}
\widetilde{B}\ Q^{-1}\ B\ \widetilde{Q}^{-1} (c^\mu\ e_\mu)\ =\
\widetilde{B}\ Q^{-1}\ B \big( c^\mu\ e_\mu \big)\ =\ c^\mu\
\widetilde{B}\ Q^{-1}\ \big(f_\mu \big)\ =\ c^\mu\ \widetilde{B}\
\big(\widetilde{D}f_\mu\big)\ =\ c^\mu \widetilde{D}f_\mu
\end{equation}
Here in the first equality we used that $\widetilde{Q}^{-1}$ is just
a choice of representative and acts as identity on the basis
elements. In the second equality we used (\ref{basis for free
resolvent}). In the third equality the definition of $Q^{-1}$ was
explored. The last equality is factorization over the ideal $I_f$.
This technique allows to find explicit expression for the cohomology
of Berkovits complex simply by the trivial application of the
zig-zag to the obvious cohomology of the vertical complex.

  This method can be straightforwardly generalized to the terms
with higher degree in $\theta_\alpha$. For  example, the zig-zag for
the class $[F(\lambda,\theta^2)]$ is given by:
$$
\begin{CD}
.....@>Q>>\ [F(\lambda,\theta^2)]\ @>Q>>\ [F(\lambda,\theta)]\ @>Q>>\ [F(\lambda)]\\
@.  @AA\widetilde{B}A @AA\widetilde{B}A @AA\widetilde{B}A @.\\
.....@>Q>>\ F(\lambda,\theta^2)\ @>Q>>\ F(\lambda,\theta)\ @>Q>>\ F(\lambda)\ @>\widetilde{Q}>>\ c\\
@.  @AABA  @AABA @AABA @AA0A @.\\
.....@>Q>>\ F^\mu(\lambda,\theta^2)e_\mu\ @>Q>>\ F^\mu(\lambda,\theta)e_\mu\ @>Q>>\ F^\mu(\lambda)e_\mu\ @>\widetilde{Q}>>\ c^\mu e_\mu\\
@.  @AABA  @AABA @AABA @AA0A @.\\
.....@>Q>>\ F^{A_1}(\lambda,\theta^2)e_{A_1}\ @>Q>>\ F^{A_1}(\lambda,\theta)e_{A_1}\ @>Q>>\ F^{A_1}(\lambda)e_{A_1}\ @>\widetilde{Q}>>\ c^Ae_A \\
@.  @AABA  @AABA @AABA @AA0A @.\\
@ ...... @ ...... @ ...... @ ......\\
\end{CD}
$$
\begin{picture}(10,10)(30,30)
\Line(190,192)(190,225)\Line(190,192)(290,192)
\Line(290,192)(290,147)\Line(290,147)(390,147)
\Line(390,147)(390,106)\Line(390,106)(450,106)
\end{picture}
Calculation similar to (\ref{first zig-zag formula}) gives:
$$
{\begin{split} \widetilde{B}\ Q^{-1}\ B\ Q^{-1}\ B\
\widetilde{Q}^{-1}\ \big(c^A e_A\big)\ =\ c^A\ \widetilde{B}\
Q^{-1}\ B\ Q^{-1}\ B\ \big( e_A\big)\ =\ c^A\ \widetilde{B}\ Q^{-1}\
B\ Q^{-1}\ \big( G_A^{(1)\
\mu}e_\mu\big)\ = \\
=\ c^A\ \widetilde{B}\ Q^{-1}\ B\  \widetilde{D}\big(G_A^{(1)\
\mu}\big) e_\mu\ =\ c^A\ \widetilde{B}\ Q^{-1}\
\widetilde{D}\big(G_A^{(1)\ \mu}\big) f_\mu\ =\  c^A\ \widetilde{B}\
\widetilde{D}\big(\widetilde{D}G_A^{(1)\ \mu} f_\mu\big)\ = \\
=\ c^A\ \widetilde{D}\big(\widetilde{D}G_A^{(1)\ \mu} f_\mu\big) \ \
\ \ \ \ \ \ \ \
\end{split}}
$$
which is the third line of (\ref{coh}).

\section{Acknowledgements}
  It is a pleasure to thank  A.~Morozov, M.~Movshev, N.~Nekrasov,
V.~Rubakov and A.~Schwarz for useful discussions. Especially we
would like to thank Alexei Gorodentsev for explaining the
geometrical meaning of the discussed complex and the role of Zig-Zag
technique. Also, we are indebted to V.Lysov for his contribution to
our understanding of Feynman diagram technique in AKSZ-like theories
and pointing out the mistake in the expressions (\ref{coh}) in the
previous version of the paper. Also we are greatly indebted to
A.~Rosly for the careful reading of the manuscript and illuminating
discussions. The work of A.L. was supported by the grants
อุ-8065.2006.2,\ \ NWO-RFBR 047.011.2004.026 (RFBR
05-02-89000-NWOa)\  and\  INTAS 03-51-6346. The work of D.K. was
supported by Dynasty Foundation and RFBR-08-02-00473.

\end{document}